\documentclass[journal=jacsat,manuscript=article]{achemso}
\usepackage[version=3]{mhchem} % Formula subscripts using \ce{}
\usepackage[T1]{fontenc}       % Use modern font encodings
\usepackage{longtable}
\usepackage{multirow} 
\usepackage{array}
\usepackage[english]{babel}

\setkeys{acs}{usetitle = true}
\usepackage{booktabs} 
\usepackage{amsmath}
\usepackage{wrapfig}
\usepackage{graphicx}
\usepackage{subfig}
\usepackage[export]{adjustbox}

\usepackage[final]{pdfpages}

\author{Marcus Djokic}   
\affiliation{Department of Chemical Engineering and Material Science, Michigan State University, East Lansing, MI, 48824, USA.}

\author{Jose L. Mendoza-Cortes}   
\email{jmendoza@msu.edu} 
\affiliation{Department of Chemical Engineering and Material Science, Michigan State University, East Lansing, MI, 48824, USA.}

\title{MultiBinding Sites United in Covalent-Organic Frameworks (MSUCOF) for \ce{H2} Storage and Delivery at Room Temperature} 

\begin{document}
	%----------------------------------------------------------------------------------------	
\begin{tocentry}

\begin{figure}[H]
{\includegraphics[height=3.5cm]{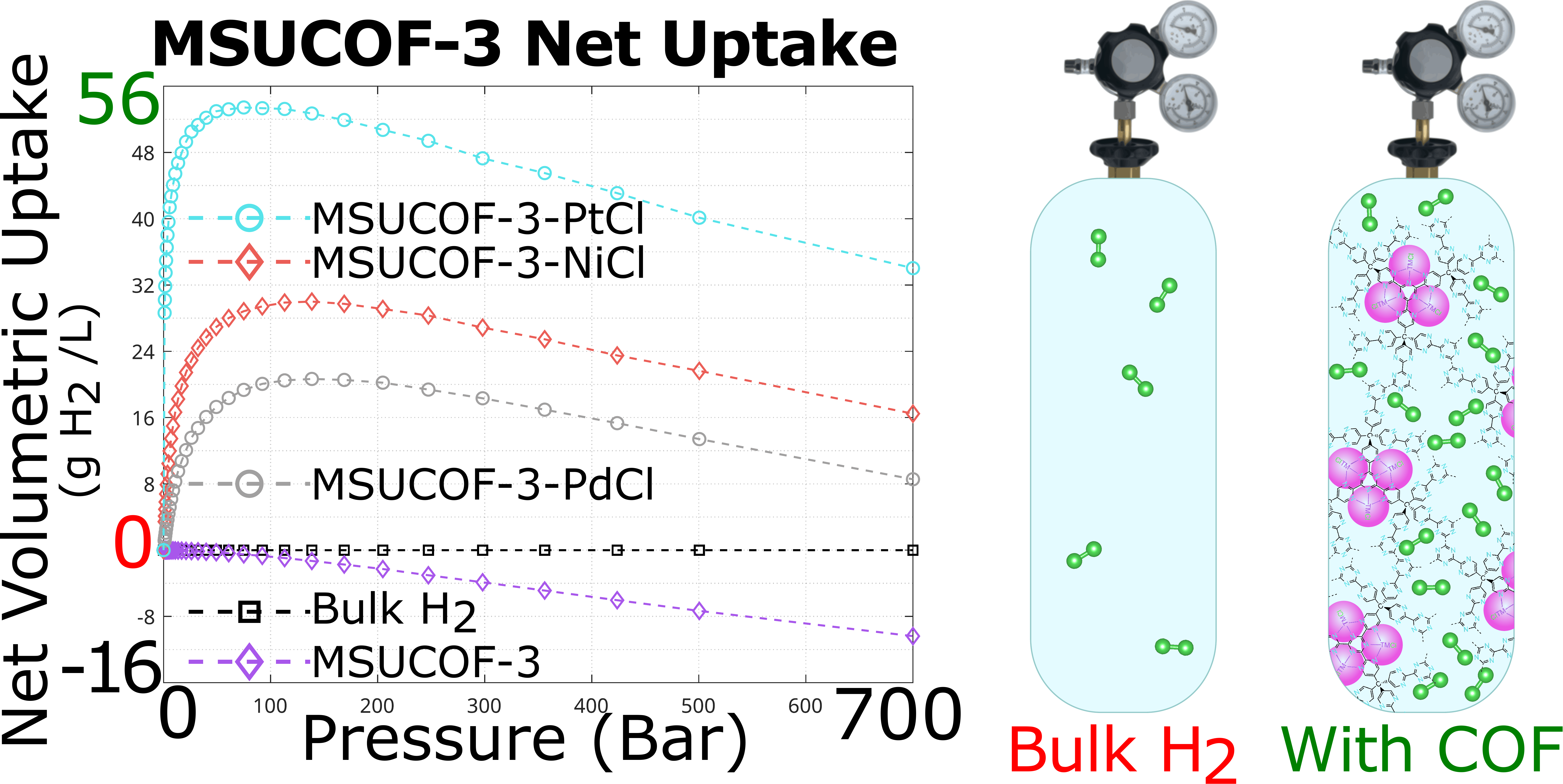}}
\label{fig:Cover_Image_TOSCALE}
\end{figure}

\end{tocentry}

	\begin{abstract}

 The storage of hydrogen gas (\ce{H2}) has presented a significant challenge that has hindered its use as a fuel source for transportation. To meet the Department of Energy's ambitious goals of achieving $50$ g L$^{-1}$ volumetric and $6.5$ wt \% gravimetric uptake targets, materials-based approaches are essential. Designing materials that can efficiently store hydrogen gas requires careful tuning of the interactions between the gaseous \ce{H2} and the surface of the material. Metal-Organic Frameworks (MOFs) and Covalent-Organic Frameworks (COFs) have emerged as promising materials due to their exceptionally high surface areas and tunable structures that can improve gas-framework interactions. However, weak binding enthalpies have limited the success of many current candidates, which fail to achieve even $10$ g L$^{-1}$ volumetric uptake at ambient temperatures. To overcome this challenge, We utilized quantum mechanical (QM) based force fields (FF) to investigate the uptake and binding enthalpies of 3 linkers chelated with 7 different transition metals (TM), including both precious metals (Pd and Pt) and first row TM (Co, Cu, Fe, Ni, Mn), to design 24 different COFs in-silico. By applying QM-based FF with grand canonical Monte Carlo (GCMC) from 0-700 bar and 298 K, We demonstrated that Co-, Ni-, Mn-, Fe-, Pd-, and Pt-based MSUCOFs can already achieve the Department of Energy's hydrogen storage targets for 2025. Surprisingly, the COFs that incorporated the more affordable and abundant first-row TM often outperformed the precious metals. This promising development brings us one step closer to realizing a hydrogen-based energy economy.

	\end{abstract}
	
	%----------------------------------------------------------------------------------------
	%	INTRODUCTION
	%----------------------------------------------------------------------------------------
	%****************************************************************************************
	\section{INTRODUCTION}
	%----------------------------------------------------------------------------------------
	%****************************************************************************************	

Storing hydrogen at operational temperatures poses a significant challenge to its commercial viability, particularly in applications such as transportation. The U.S. Department of Energy (DOE) has set ambitious targets for hydrogen storage by 2025, including achieving 5.5 weight percent (wt \%) and 40 g L$^{-1}$ at 233-358 K, with an ultimate target of 6.5 wt \% and 50 g L$^{-1}$.\cite{doe_doe_2017} Additionally, during the DOE's Hydrogen Shot summit in 2021, a further goal was announced: to reduce the cost of clean hydrogen to \$1 per 1 kg within the next decade.\cite{dinh_hydrogen_2022} To achieve these targets, it is evident that materials-based approaches are preferred. Rather than attempting to compress or cryogenically cool hydrogen storage tanks, adsorbents such as Metal-Organic Frameworks (MOFs) and Covalent-Organic Frameworks (COFs) can be introduced to the vessel, improving gas uptakes without incurring high utility costs. Of these adsorbents, COFs show particular promise due to their high surface area and composition of lighter, more abundant elements compared to MOFs. \cite{han_recent_2009, geng_covalent_2020, furukawa_storage_2009}

While many MOFs can offer high uptakes, operating at cryogenic temperature (77 K) swings to ambient conditions (298 K) is often prohibitively expensive. To make the process more economical, it is preferable to perform the adsorption process at ambient conditions. However, this is challenging for many MOFs and COFs, as adsorption energies decay at higher temperatures due to entropic changes. \cite{bhatia_optimum_2006,han_high_2008} As a result, MOFs and COFs can lose their loading capacity at higher temperatures, and those with promising uptakes at 77 K may not perform as well at room temperature.

Efforts to improve gas uptake in porous materials such as MOFs and COFs have led to the development of several strategies. One approach is reticular chemistry, which allows for the selection of different nodes and linkers to create customized structures. Building on this concept, Han et al. investigated how various factors, including pore size, impregnation, catenation, open metal sites, and functionalized linkers, affect hydrogen storage in MOFs and COFs.\cite{han_recent_2009} In our study, we demonstrate that COFs with smaller pores and tri-topic metalated linkers exhibit state-of-the-art performance. To achieve optimal results, it is essential that metal sites are easily accessible, enabling them to interact more efficiently with adsorbates. We propose that by incorporating multiple metal sites into a linker, a greater number of \ce{H2} molecules can interact directly with the sites via long-range interactions without being shielded. Furthermore, cooperative interactions may occur if these metalated sites are in close proximity. By exploring these ideas, we aim to advance the field's understanding of how customizable structures and metalated linkers can enhance gas uptake in porous materials.

The conventional approach for predicting gas uptake in COFs and MOFs at cryogenic temperatures (77 K) relies on pore volume, void fraction, and surface area.\cite{ahmed_exceptional_2019,ahmed_predicting_2021} However, these features overlook crucial gas-framework interactions, such as non-covalent and orbital interactions. To achieve optimal hydrogen storage at room temperature and pressures up to 700 bar, it is crucial to consider binding energies ($\Delta H_{bind}^0$) in the range of 7-15 kJ/mol, which fall between chemi- and physisorption.\cite{kubas_fundamentals_2007, pramudya_design_2016, bhatia_optimum_2006, han_recent_2009} Since COFs typically consist of many repeating binding sites, $\Delta H_{bind}^0$ is a good approximation to the experimentally obtained initial isosteric heat of adsorption (Q$_{st}^0$). Therefore, this heuristic of 7-15 kJ/mol should also apply to the Q$_{st}^0$. The results in this paper further support this correlation that the 7-15 kJ/mol Q$_{st}^0$ range results in optimal performance.

%----------------------------------------------------------------------------------------
%	MATERIALS AND METHODS
%----------------------------------------------------------------------------------------
\section{MATERIALS AND METHODS}
%----------------------------------------------------------------------------------------

\subsection{Design of Triazine COFs}

\begin{figure}[htp!]
        \includegraphics[width=1\linewidth]{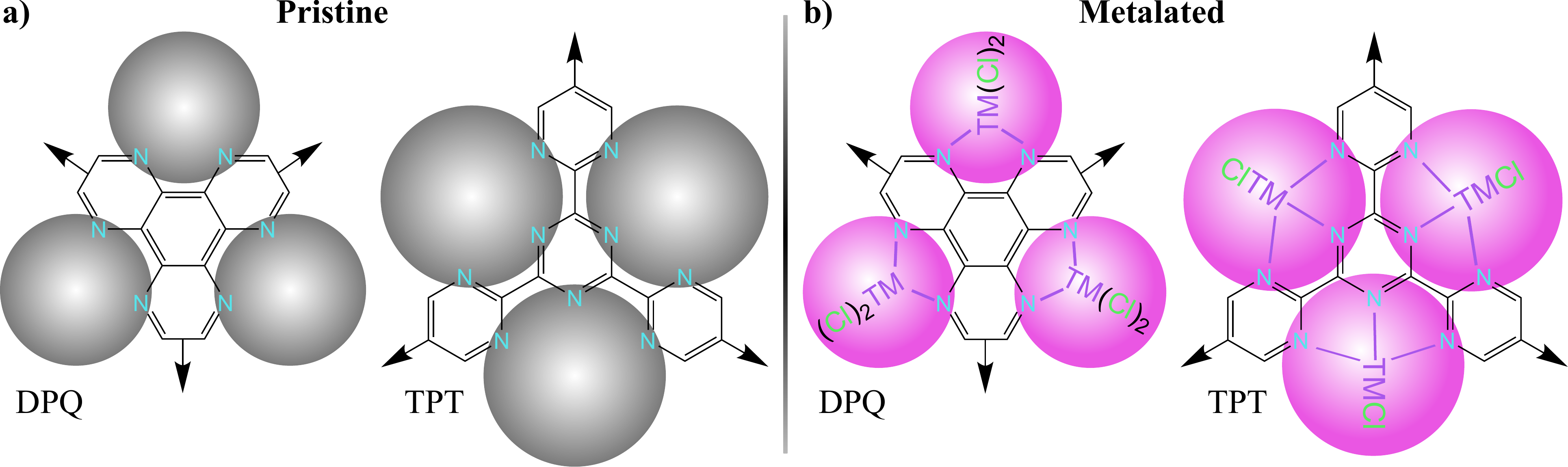}
	\caption{New linkers with tri-topic connectivity for N-containing-COFs where a high density of metalation sites are present: Dipyrazino [2,3-f:2',3'-h] quinoxaline (\textbf{DPQ}) and 2,4,6-tri(pyrimidin-2-yl)-1,3,5-triazine (\textbf{TPT}). The a) pristine linkers are shown compared to b) the metalated case. Note that \textbf{DPQ} sites can accommodate TM\ce{Cl2}, while \textbf{TPT} sites accommodate TM\ce{Cl}.}
	\label{fig:Linker}
\end{figure}

To create new frameworks with a high density of metalation sites and cooperative effect, we introduced two linkers (Figure \ref{fig:Linker}): Dipyrazino [2,3-f:2',3'-h] quinoxaline (\textbf{DPQ}) and 2,4,6-tri(pyrimidin-2-yl)-1,3,5-triazine (\textbf{TPT}). These linkers were inspired by the well-known 4,4'-bipyridine (\textbf{BPY}) and 4,4'-diethylamido-2,2'-bipyridine (\textbf{BPYAM}) linkers and were chosen to increase the density of metalation sites. To generate triazine-COFs, we constrained the connectivity to a semi-regular net with a 3-4 connection. We selected only the \textbf{ctn} ($I-43d$ space group) topologies for these frameworks, as they have been shown to be the most stable.\cite{geng_covalent_2020,el-kaderi_designed_2007, delgado-friedrichs_three-periodic_2006, schmid_accurate_2008} We minimized the frameworks using molecular mechanics without any symmetry constraints, and the coordinates are reported in the SI. We note that related triazine frameworks have been reported experimentally.\cite{kuhn_porous_2008, singh_rational_2022,sun_three-dimensional_2022,lin_covalent_2022}

\begin{figure}[htbp!]
	%	\vspace{3mm}
	\includegraphics[width=1\linewidth]{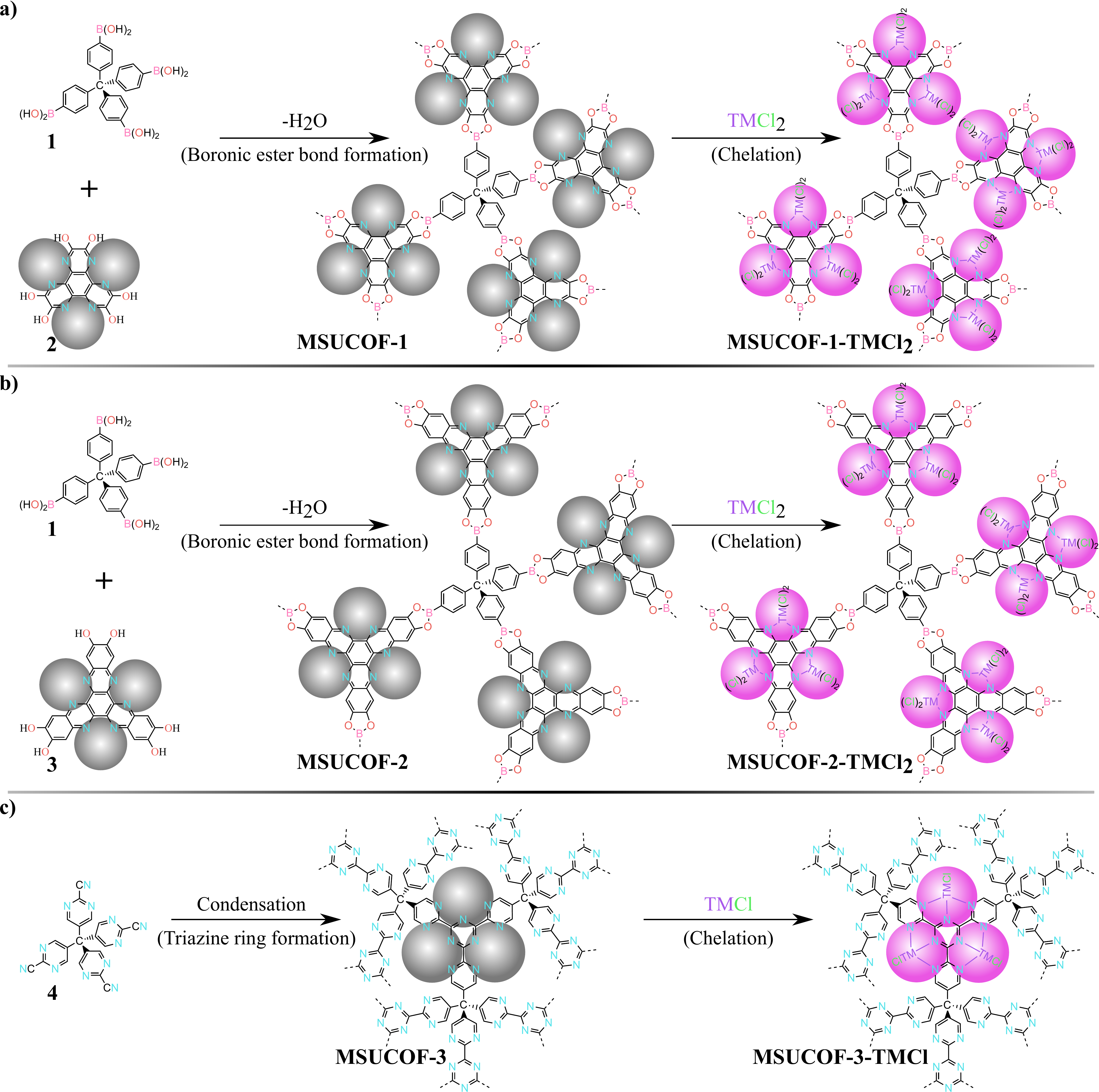}
	%	\vspace{5mm}
	\caption{Approach suggested for the incorporation of the hydrazine functionality with known boroxine and chelation chemistry to form a) MSUCOF-1 and MSUCOF-1-TM\ce{Cl2}, b) MSUCOF-2 and MSUCOF-2-TM\ce{Cl2}, or by in-situ triazine formation to form c) MSUCOF-3 and MSUCOF-1-TM\ce{Cl}}
	\label{fig:formation}
	%	\vspace{-10mm}  
\end{figure}

\begin{figure}[htp!]
	%	\vspace{3mm}
	\includegraphics[width=1\linewidth]{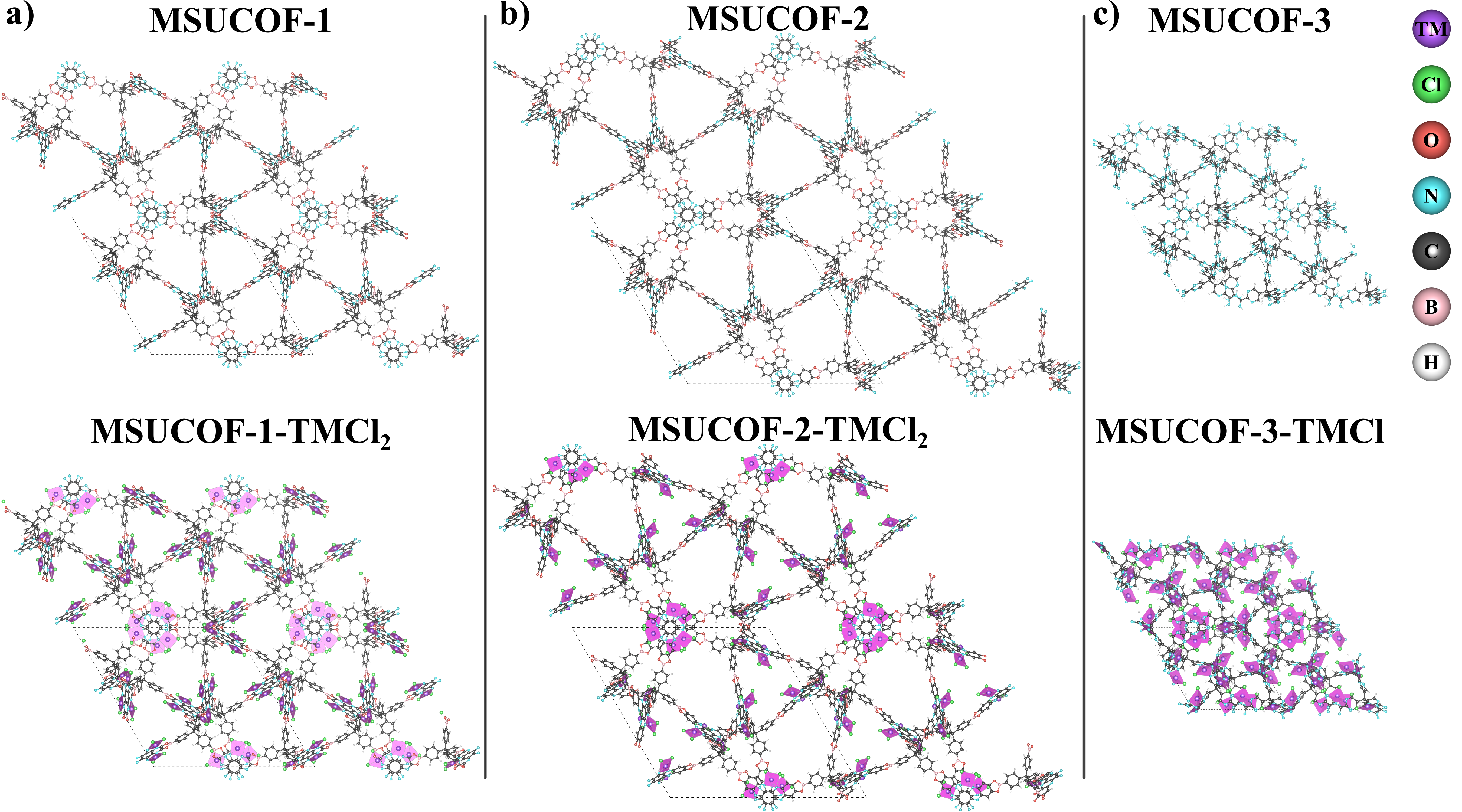}
	%	\vspace{5mm}
	\caption{FF-Optimized structures of the pristine and metalated cases for a) MSUCOF-1, b) MSUCOF-2, and c) MSUCOF-3.}
	\label{fig:VESTA}
	%	\vspace{-10mm}  
\end{figure}

To generate new N-containing COFs, we utilized four experimental building blocks (\textbf{1}-\textbf{4}) as shown in Figure \ref{fig:formation}.
These include methanetetrayltetrakis (benzene-4,1-diyl) tetraboronic acid (\textbf{1}), dipyrazino [2,3-f:2',3'-h] quinoxaline-2,3,6,7,10,11-hexaol (\textbf{2}), diquinoxalino [2,3-a:2',3'-c] phenazine-2,3,8,9,14,15-hexaol (\textbf{3}), and 5,5',5'',5'''-methanetetrayl-tetrakis (pyrimidine-2-carbonitrile) (\textbf{4}).  We used the known boronic chemistry to create MSUCOF-1 (\textbf{1}+\textbf{2}) and MSUCOF-2 (\textbf{1}+\textbf{3}). \textbf{3}  serves as an extended version of \textbf{2}  with the addition of benzene rings to each tri-topic end of \textbf{2}, in effect creating a MSUCOF-2 with a larger pore size, \textit{P$_{size}$}, than MSUCOF-1 (as seen in Figure \ref{fig:VESTA}). Linker \textbf{1} has already been used for imine COFs.\cite{waller_chemistry_2015, geng_covalent_2020}  By using the new triazine formation method, we design MSUCOF-3 (self-condensation of \textbf{4}). Ultimately, MSUCOF-1 and MSUCOF-2 have  \textbf{DPQ} sites, while MSUCOF-3 has \textbf{TPT} sites. 
Upon metal chelation, TM-Cl$_\text{x}$ (where TM = Co, Cu, Fe, Mn, Ni, Pd, and Pt) will occupy the DPQ and TPT sites to form metalated COFs using known methods. The intention is to form interaction between the s-orbitals from \ce{H2} and the d-orbitals from the TM.\cite{kubas_fundamentals_2007}

\subsection{Volumetric and Gravimetric uptake calculation}

When analyzing porous materials, there are multiple methods to compare their properties. Gas absorption is typically assessed by calculating volumetric uptakes (g of \ce{H2} L$^{-1}$) and gravimetric uptakes (g of \ce{H2} g$^{-1}$). Gravimetric uptakes are particularly important in vehicles since they measure the added weight contribution of the porous material. Volumetric uptakes, on the other hand, evaluate the energy density of hydrogen storage relative to the container size. To accurately evaluate gas adsorption, it is essential to distinguish between absolute, excess, and net adsorption. Absolute uptakes measure the total number of \ce{H2} molecules loaded into the structure, whereas excess uptakes quantify the loading with respect to \ce{H2} that would occupy the pore volume if there were no COF/MOF present, as shown in Equation \ref{eq:excessuptake}. \cite{brandani_net_2016}

\begin{equation} \label{eq:excessuptake}
\centering
n^{exc.} = n^{abs.} - V_P * \rho^{exp.}_{H_{2}}
\end{equation}

Here, $n^{exc.}$ is the quantity of excess gas molecules, $n^{abs.}$ is the absolute total quantity of gas molecules in the adsorbent system, $V_p$ is the pore volume, and $\rho^{exp.}_{H_{2}}$ is the experimental hydrogen density at the same temperature and pressure as the adsorbent system. In our calculations, we used the NIST database to obtain hydrogen gas densities.\cite{linstrom_nist_2001} A major pitfall to excess uptakes is it fails to take into account the space contribution of the adsorbent itself, which makes the adsorbent appear to be more promising. A positive excess uptake may still result in a worse performance than an tank full of hydrogen alone at the same pressure. Additionally, excess uptake is directly sensitive to how the pore volume is determined (e.g. geometric pore volume, probe-occupiable pore volume, van der Waals volume, Connolly volume, and more)\cite{ongari_accurate_2017,herrera_self-consistent_2012,connolly_computation_1985}, which pores are accessible\cite{van_heest_identification_2012}, and whether the pore volume can change as is the case with flexible MOFs/COFs \cite{schneemann_flexible_2014,sakamoto_unusual_2022}. Net uptakes, shown in Equation \ref{eq:netuptake}, try to address these issues by measuring uptake with respect to the empty unit cell at the same pressure, thereby directly assessing whether the adsorbent improves overall storage capacity. 

\begin{equation} \label{eq:netuptake}
\centering
n^{net} = n^{abs.} - V_{U.C.} * \rho^{exp.}_{H_{2}}
\end{equation}

Replacing $V_p$ in Equation \ref{eq:excessuptake} with the unit cell volume of the MOF/COF, denoted as $V_{U.C.}$, yields the net uptake, $n^{net}$. A positive net uptake indicates an improvement in gaseous loading when the adsorbent is added to a storage vessel. Nevertheless, net, absolute, and excess uptakes do not consider the kinetics of the adsorption and desorption processes. An adsorbent with high uptakes may not be practical if it cannot release hydrogen quickly and inexpensively. To address this, usable working capacity (\textit{WC}) quantifies the amount of gas that can be withdrawn from the adsorbent in a reasonable time frame. Specifically, \textit{WC} can be calculated by subtracting the absolute uptake at 5 bar, $n^{abs.}_{5}$, from the maximum uptake at the highest pressure tested, $n^{abs.}_{700}$ (in this study, 700 bar).

\begin{equation} \label{eq:wc}
\centering
WC = n^{abs.}_{700} - n^{abs.}_{5} 
\end{equation}

\subsection{QM-Based Force Fields}

To accurately capture the interaction between molecular hydrogen and transition metal chelation complexes (framework-TMCl$_\text{x}$), a Morse potential (shown in Equation \ref{eq:Morse}) was fitted and used to parameterize the interaction. The fitting process involved considering various first-row elements (Co, Cu, Fe, Mn, and Ni) and late precious metals (Pd and Pt) using density functional theory (DFT). The first row TM were previously fit\cite{pramudya_design_2016} using B3LYP-D3 in Amsterdam Density Functional and GULP. In this paper, this same methodology was used to expand to Pd and Pt, where the Morse fitted parameters are summarized in Table \ref{Morse}. These parameters were incorporated as non-bonding terms in the QM-modified force field, allowing for more accurate modeling of the porous framework and gas interaction. It is important to note that the Morse potential is specifically tailored to the adsorption of molecular hydrogen and the transition metal chelation complexes in the framework, ensuring a more precise representation of the system.

\begin{equation} \label{eq:Morse}
\centering
U_{ij}^{\text{Morse}} (r_{ij}) = D_0[(1-\text{e}^{-\alpha(r_{ij}-r_{0})})^2-1] 
\end{equation}

\begin{table}[h!]
	\centering
	\small
	\begin{tabular}{lccc} 
            \hline
		\textbf{Element}	&\textbf{D$_{\mathbf{0}}$ (kcal/mol)}	&\textbf{$\pmb{\alpha}$ (\AA$^{-1}$)}		&\textbf{r$_{\mathbf{0}}$ (\AA)} \\
            \hline
            \ce{H2}-Co  &0.879  &0.850  &2.985\\ 
		\ce{H2}-Cu	&0.818	&1.462 	&2.931\\
		\ce{H2}-Fe	&1.092	&1.180 	&3.155\\
		\ce{H2}-Mn	&0.994	&0.990 	&3.015\\
		\ce{H2}-Ni	&1.154	&1.210 	&3.207\\
		\ce{H2}-Pd	&0.641 	&0.974 	&3.268\\
		\ce{H2}-Pt	&1.652 	&0.920 	&3.282\\
            \ce{H2}-Cl	&0.146	&1.446 	&3.725\\
		\end{tabular}
	\caption{Table of fitted Morse potential force field parameters between chelated atoms and \ce{H2}}\label{Morse}
\end{table}

\doublespacing 

\subsection{Grand Canonical Monte Carlo (GCMC)}
 
In this study, we implemented the Ab-initio parameterized Force Fields in Grand Canonical Monte Carlo (GCMC) simulations. To perform these simulations, we employed the Metropolis algorithm code implemented in the Sorption Module of Materials Studio\cite{biovia_materials_2022}. Our investigations were conducted at 298 K and pressures ranging from 1 bar to 700 bar. During each simulation step, hydrogen gas underwent various processes in a predefined ratio (2:1:1:1:1:0.1) for translation, rotation, creation, annihilation, regrowth, respectively. The translation and rotation steps allowed the hydrogen atoms to freely move within the rigid COF pores, while the creation and annihilation steps facilitated the addition and removal of hydrogen molecules. The regrowth process accommodated the conformational changes in asymmetric molecules, which for molecular hydrogen, resulted in translation or rotation. Equilibrium was achieved after 1,000,000 simulation steps, and the following 3,000,000 steps were utilized to calculate the ensemble average loading and isosteric heat of adsorption (Q$_{st}$) for each temperature and pressure point.

\subsection{Equation of State}

\begin{gather*}
P  =\frac{RT}{\nu-b}-\frac{a}{\nu^2} 
\end{gather*}
\begin{equation} \label{eq:vdw}
 ln \frac{f}{P} = (b-\frac{a}{RT})\frac{P}{RT} 
\end{equation}

\begin{figure}[htp!]
	\includegraphics[width=0.60\linewidth]{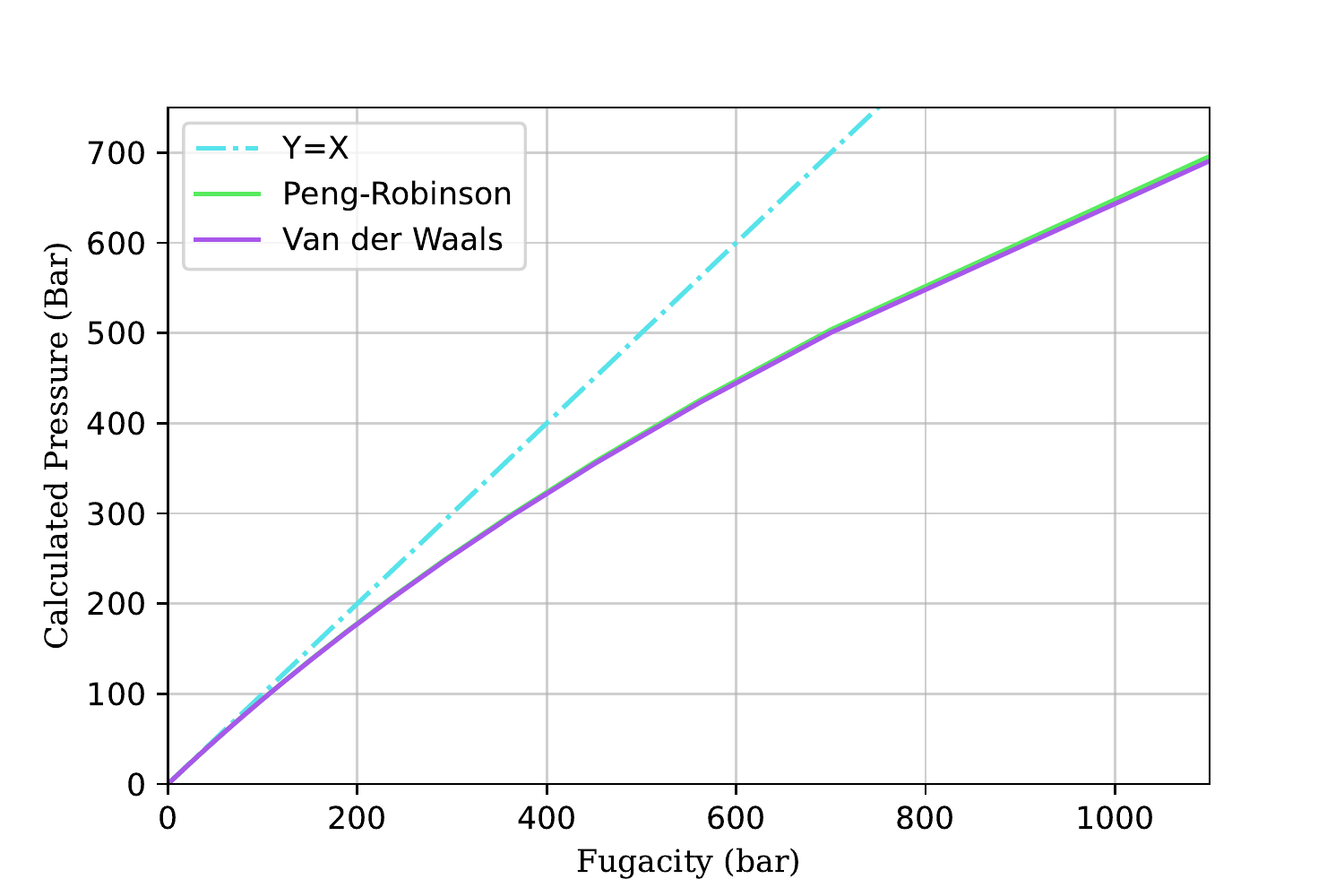}
	\caption{Calculated Van der Waals (purple) and Peng-Robinson (green) fugacity to pressure conversion showing good agreement up to the 700 bar pressure evaluated. It is clear that these pressure and fugacity conversion stray from the ideal gas approximation (light blue y=x line) above values of ~100 bar for \ce{H2} gas at 298 K.}
	\label{fig:EOS}
\end{figure}

To make these results applicable to real-world data, an equation of state (EOS) is required to convert the computed fugacity to pressure. The Van der Waals\cite{winn_fugacity_1988}(Equation \ref{eq:vdw}) and Peng-Robinson\cite{peng_new_1976} (Equation \ref{eq:peng}) equations of state (EOS) were both considered. In each equation, empirical terms $a$ and $b$ are used, with values of 0.2476 L$^2$ bar/mol$^2$ and 0.02661 L/mol, respectively, for hydrogen. Other variables include $T$ for temperature, $\nu$ for molar volume, $R$ for the gas constant, $P$ for pressure, $f$ for fugacity, and $Z$ for compressibility factor. Note that for simplicity, terms A, B, and C are used which represent various groupings of the previously mentioned variables into single terms. Figure \ref{fig:EOS} demonstrates the differences between an ideal gas (y=x), Van der Waals EOS (blue), and the Peng-Robinson EOS (orange). In the low pressure regime <100 bar, Hydrogen gas will act like an ideal gas. At the 700 pressure range however, it strays drastically from the ideal pressures.  Between the two EOS, the calculated pressures showed a 0.735\% difference between the two EOS. Though both seem to be fairly interchangeable under these conditions, we chose to use the Van der Waals EOS for fugacity and pressure conversions.

\begin{gather*}
P  =\frac{RT}{\nu-b}-\frac{a(T)}{\nu(\nu+b)+b(\nu-b)} \\
Z^3 - (1-B)Z^2 + (A - 3B^2 - 2B)Z - (AB - B^2 - B^3) = 0 \\
A  =\frac{aP}{R^2T^2} , B =\frac{bP}{RT}, C = \frac{P\nu}{RT} 
\end{gather*}
\begin{equation} \label{eq:peng}
 ln \frac{f}{P} = Z-1-ln(Z-B)-\frac{A}{2\sqrt{2}B}ln(\frac{Z+2.414B}{Z-0.414B})  
\end{equation}

	%****************************************************************************************

%----------------------------------------------------------------------------------------
%	RESULTS AND DISCUSSION
%----------------------------------------------------------------------------------------
\section{RESULTS AND DISCUSSION}
%----------------------------------------------------------------------------------------

\begin{wrapfigure}{l}{0.51\textwidth}
    \includegraphics[width=0.51\textwidth]{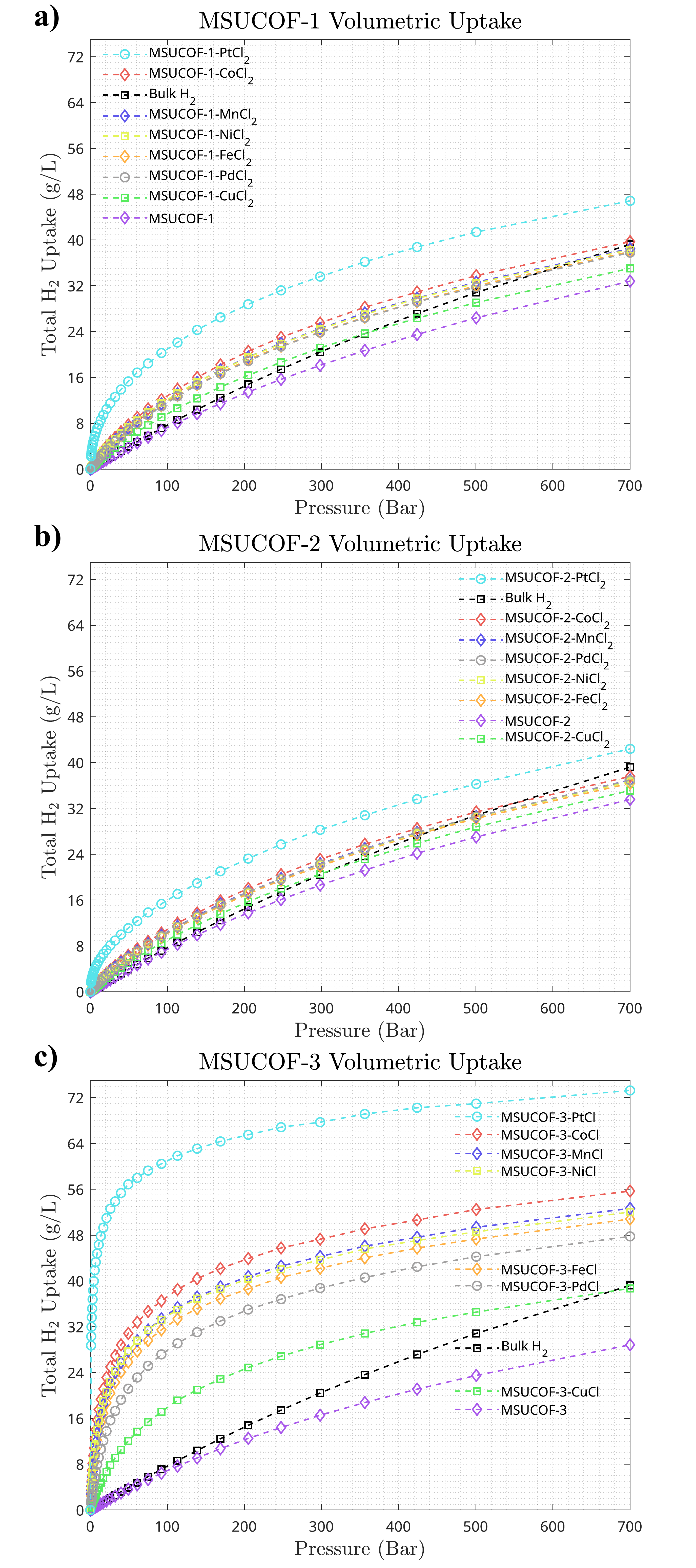}
    \caption{We show the high pressure \ce{H2} volumetric isotherms at 298 K for (a) MSUCOF-1-TM\ce{Cl2}, (b) MSUCOF-2-TM\ce{Cl2}, and (c) MSUCOF-3-TMCl.}
    \label{fig:vu}
\end{wrapfigure}

The \ce{H2} uptakes for the new N-triazine-COFs were calculated from 1 to 700 bar at 298 K (Figure \ref{fig:vu}). The un-metalated COFs of this family give total volumetric uptakes of 28 to 33 g L$^{-1}$ while the metalated cases give uptakes of at least 35 g L$^{-1}$ at 700 bar. Of the three COF families tested, the MSUCOF-3 family gives the best performance in volumetric units of up to 73 g L$^{-1}$, where MSUCOF-3-Pt\ce{Cl2} obtains approximately triple the uptake of the pristine MSUCOF-3. This is due to the high interaction of the MSUCOF-3 framework with \ce{H2} while also having the smallest \textit{P$_{size}$} of 7.7-9.4 \AA,  which avoids wasting useful space. Congruently, this is close to the postulated range of the ideal \textit{P$_{size}$} for \ce{H2} storage of 6 to 12 \AA. Thus these new COFs can overcome the 2025 DOE volumetric target of 40 g/L and gravimetric target of 5.5 weight \%. The excess gravimetric uptakes of the pure N-triazine-COFs are very similar to other top-performing COFs. The metalated triazine-COFs perform as well as the previously discovered promising COF-301-Pd\ce{Cl2} with total 60 g L$^{-1}$ and 4.2 excess wt \% uptakes at 100 bar and 298 K.\cite{mendoza-cortes_covalent_2012} In comparison, MSUCOF-3-Pt\ce{Cl2} at 100 bar and 298 K achieve a total uptake of 60.4 g L$^{-1}$ and 4.06 excess wt \%.

\newpage

\begin{wrapfigure}{R}{0.51\textwidth}
	\includegraphics[width=0.51\textwidth]{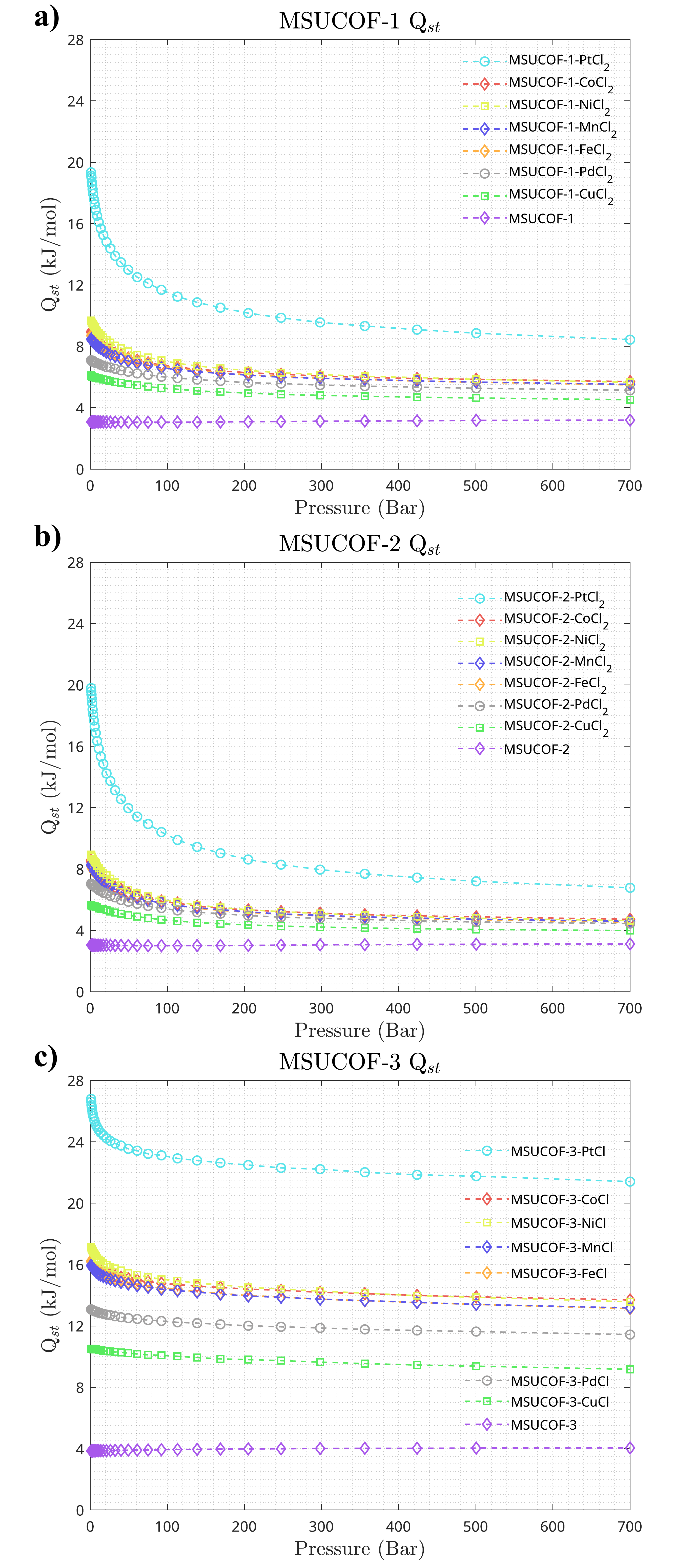}
	\caption{We show the high pressure \ce{H2} isosteric heat of adsorption, Q$_{st}$, at 298 K for (a) MSUCOF-1-TM\ce{Cl2}, (b) MSUCOF-2-TM\ce{Cl2}, and (c) MSUCOF-3-TMCl.}
	\label{fig:qst}
\end{wrapfigure}

For the unmetalated triazine-COFs, we calculated the isosteric heat of adsorption (Q$_{st}$) to be around 3-4 kJ mol$^{-1}$, as shown in Figure \ref{fig:qst}. However, upon metalation, the Q$_{st}$ values increased significantly, ranging from 4-26 kJ mol$^{-1}$. Notably, the MSUCOF-3 family exhibited the highest Q$_{st}$ values, likely due to its small pore size and multiple binding sites per pore. This trend is reminiscent of the COF-301-Pd\ce{Cl2} case, where the pore size (\textit{P$_{size}$}) was 6.3 \text{\AA} and the space between neighboring framework layers (\textit{D$_{layers}$}) was 9.9 \text{\AA}. We propose that the compact nature and multiple binding sites of MSUCOF-3 may induce a cooperative effect, resulting from the overlap of the energy surfaces of adjacent metalated sites.

A common trend observed in Q$_{st}$ values is that the maximum values are achieved at lower pressures. As gas molecules are introduced into the structure at higher pressures, the interactions between the framework and \ce{H2} become obscured by the presence of other \ce{H2} molecules in the pore. The initial few layers of adsorption can only accommodate a limited number of molecules, 

\noindent while the bulk pore can efficiently incorporate more hydrogen molecules. As a result, the initial Q$_{st}$ value, Q$_{st}^0$, at 1 bar gradually transitions to a predominantly bulk-filled interaction as the pressure increases up to 700 bar (refer to Table \ref{summary}).

To investigate whether other transition metals could perform similarly to Pd or Pt, we calculated the \ce{H2} uptake for MSUCOF-1-TM\ce{Cl2}, MSUCOF-2-TM\ce{Cl2}, and MSUCOF-3-TM\ce{Cl2}, where TM includes Co, Ni, Fe, Cu, Mn, Pd, and Pt. Interestingly, several of the tested early transition metals exhibited high uptakes that were comparable to those of Pd.

\begin{figure}[htp!]
	\includegraphics[width=0.51\linewidth]{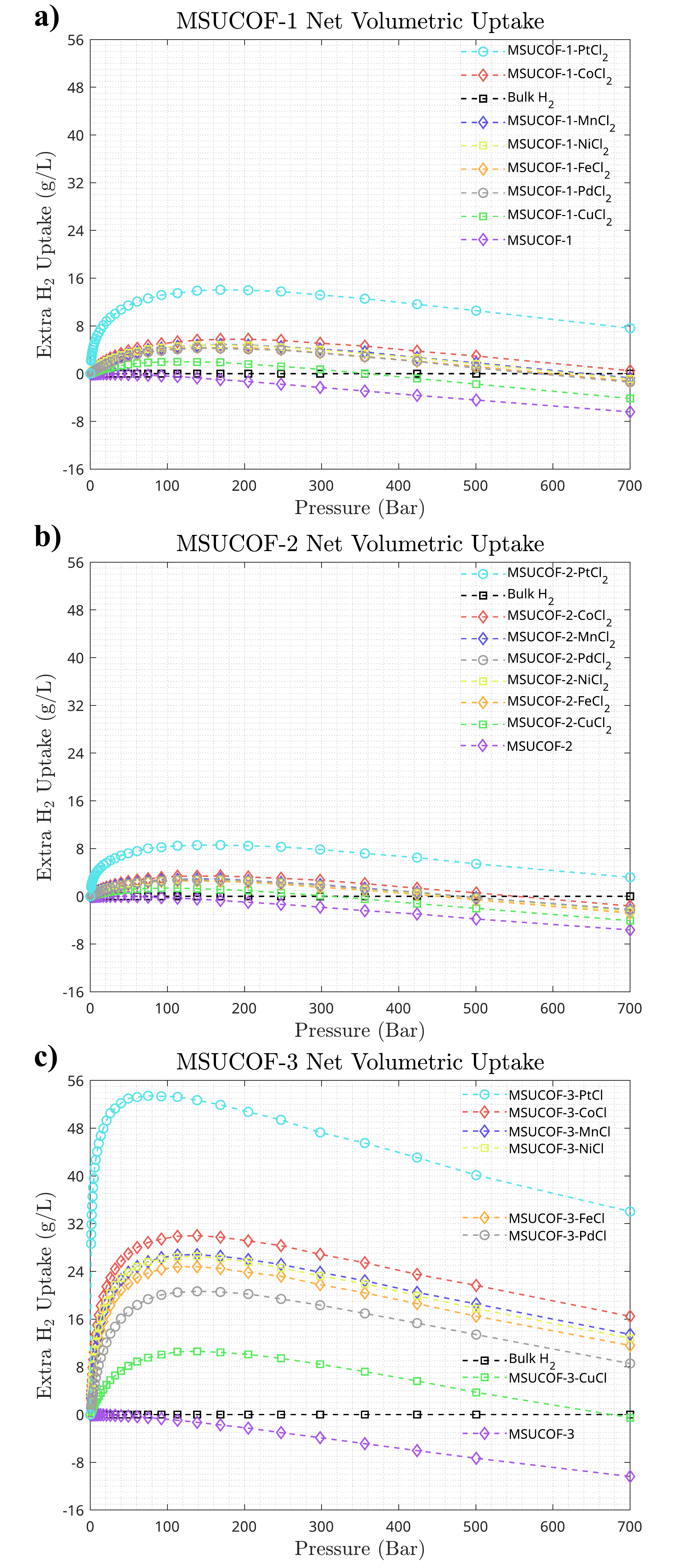}
	\caption{We show the high pressure \ce{H2} net volumetric isotherms, which compares uptakes with respect to \ce{H2} loading in an empty unit cell of the same volume, at 298 K for (a) MSUCOF-1-TM\ce{Cl2}, (b) MSUCOF-2-TM\ce{Cl2}, and (c) MSUCOF-3-TMCl.}
	\label{fig:nvu}  
\end{figure}

Among the early transition metals tested, Co, Ni, Mn, and Fe showed the best performance across the new MSUCOF families. Notably, MSUCOF-3-CoCl exhibited the highest volumetric uptake of 55.7 g L$^{-1}$ at 700 bar, followed by MSUCOF-3-PdCl with 47.8 g L$^{-1}$, MSUCOF-1-Co\ce{Cl2} with 39.7 g L$^{-1}$, and MSUCOF-1-Pd\ce{Cl2} with 37.8 g L$^{-1}$. These results demonstrate that early transition metals can achieve comparable performance to heavy transition metals, which is particularly significant because 

\begin{wrapfigure}{r}{0.51\textwidth}
	\includegraphics[width=0.51\textwidth]{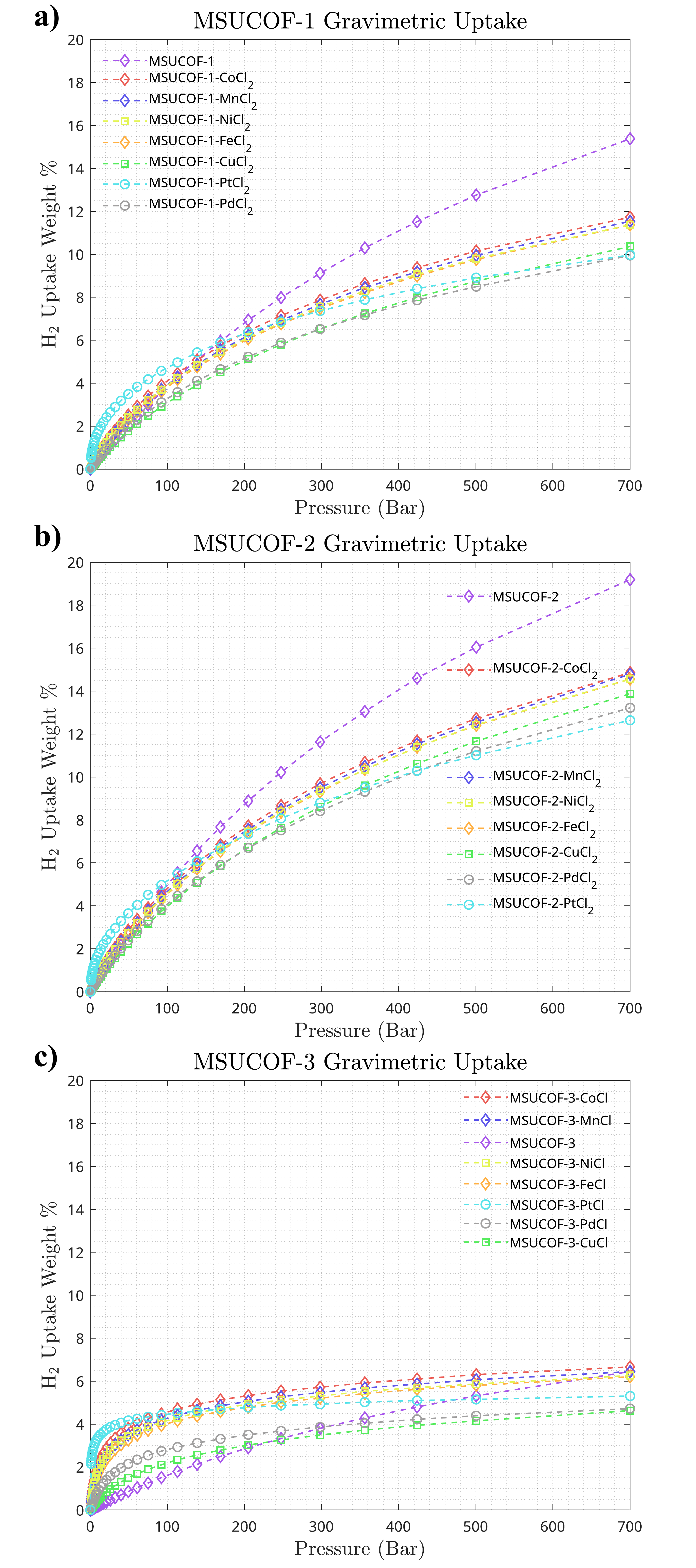}
	\caption{We show the high pressure \ce{H2} gravimetric isotherms at 298 K for (a) MSUCOF-1-TM\ce{Cl2}, (b) MSUCOF-2-TM\ce{Cl2}, and (c) MSUCOF-3-TMCl.}
	\label{fig:gu}
\end{wrapfigure}

\noindent early transition metals are relatively abundant and less expensive. In particular, \ce{CoCl2}, \ce{NiCl2}, and \ce{FeCl2} chelation reactants can be obtained at prices (\$ mol$^{-1}$) that are one to two orders of magnitude cheaper than \ce{PtCl2} or \ce{PdCl2} (see SI Section Economic Analysis of MSUCOFs). Additionally, \ce{MnCl2} and \ce{CuCl2} can be purchased at prices three orders of magnitude cheaper than the precious transition metal chlorides, which is even more impressive.

To quantify the superiority of the COFs over an empty \ce{H2} tank, it is essential to calculate the net volumetric uptakes. A positive value of net uptake indicates that the COF can accommodate more hydrogen in the storage vessel compared to bulk \ce{H2} at the same pressure. The peaks on the net volumetric chart reveal the optimal pressure (\textit{P$_{opt}$}) at which the COF performs best. Across the three families of COFs, the \textit{P$_{opt}$} occurs at approximately 100 bar, suggesting that the performance diminishes at higher pressures. This is likely due to a higher proportion of \ce{H2} bulk pore filling as adsorption sites begin to saturate. Yet as illustrated in figure \ref{fig:nvu}, MSUCOF-3-TMCl (TM = Pt, Co, Mn, Ni, Fe, Pd) demonstrates extra \ce{H2} uptakes of 8.55-34.0 g L$^{-1}$ at 700 bar, which implies that these highlighted COFs remain exceptional even in applications up to 700 bar.

Improving volumetric uptakes often comes at the expense of gravimetric uptakes, as seen in many strategies. For example, COFs with smaller pores, such as MSUCOF-3, tend to be denser than those with larger pores, such as MSUCOF-1 and MSUCOF-2. Although the MSUCOF-3 family exhibits the most impressive volumetric uptakes among these three, it also has the lowest gravimetric uptake of up to 6.7 wt \% (Figure \ref{fig:gu}). In contrast, the MSUCOF-2 family boasts an exceptional gravimetric uptake of up to 19.2 wt \%. The incorporation of TM through chelation significantly increases volumetric at the expense of slightly diminished gravimetric uptakes. Heavier precious metals add significant weight to the COFs with similar volumetric performance as first row TM. Therefore, pristine COFs, as well as chelated COFs with the most promising first-row TM, achieve the maximum gravimetric uptakes. This further demonstrates how Co, Ni, Mn, and Fe may be more favorable than Pd and Pt, as they often reach similar uptakes while contributing much less weight and cost.

Table \ref{summary} reveals some interesting trends. Firstly, metalation of COFs often resulted in up to 50\% improvement in volumetric working capacities compared to the pristine COFs. However, the addition of chelation complexes increased the weight of COFs, resulting in a reduced gravimetric working capacity. Secondly, COFs with Q$_{st}^0$ values in the 7-15 kJ/mol range tend to exhibit the best \textit{WC}. For instance, although MSUCOF-3-PtCl demonstrated impressive absolute and net uptakes, its \textit{WC} was inferior to that of the three COF families tested, owing to an overly strong interaction with hydrogen gas, which leads to a large number of loaded molecules strongly binding to the COF, even at low pressures. Third, the Q$_{st}$ tends to decrease to a bulk-filling interaction as gas is loaded into the COF. COFs that maintain Q$_{st}$ within this ideal range even at 700 bar exhibit the most impressive \textit{WC}. Although MSUCOF-1-Pt\ce{Cl2} achieved DOE's gravimetric and volumetric targets, several COFs in the MSUCOF-3 family achieved higher volumetric \textit{WC}. Importantly, common features like \textit{S$_\text{A}$} and \textit{V$_\text{P}$} had little to no bearing on predicting which COFs would perform the best.

\begin{table}
	\centering
        \small
        \tabcolsep=0.1cm
	\begin{tabular}{lcccccccc} 
            \hline
			&	&	&    & \textbf{1 bar}    & \textbf{700 bar}  &   &\multicolumn{2}{c}{\textbf{\textit{WC} (700-5)}}\\
            \hline
            \textbf{COF} &\textit{\textbf{S$_\mathbf{A}$}} &\textit{\textbf{V$_\mathbf{P}$}} &\textit{\textbf{P$_{\mathbf{size}}$}}   &\textbf{Q$_{\mathbf{st}}^\mathbf{0}$}  &\textbf{Q$_{\mathbf{st}}$} &\textbf{\textit{P$_{\mathbf{opt}}$}}   &\textbf{\ce{H2} Upt.} &\textbf{\ce{H2} Upt.} \\
                &\textbf{(m$^\mathbf{2}$/g)}   &\textbf{(cm$^\mathbf{3}$/g)}  &\textbf{(\AA)}   &\textbf{(kJ/mol)}    &\textbf{(kJ/mol)}  &\textbf{(bar)} &\textbf{(g/L)} &\textbf{(wt \%)} \\
            \hline
            MSUCOF-1               &5060   &5.06   &18.5  &3.08   &3.19   &-      &32.4   &15.2\\ 
		MSUCOF-1-Co\ce{Cl2}    &4330   &2.97   &15.2  &8.97   &5.70   &205    &38.6   &11.4\\    
		MSUCOF-1-Cu\ce{Cl2}    &4060   &2.92   &16.8  &6.10   &4.52   &113    &34.4   &10.1\\
            MSUCOF-1-Fe\ce{Cl2}    &4400   &2.99   &15.2  &8.70   &5.49   &139    &37.0   &11.0\\
            MSUCOF-1-Mn\ce{Cl2}    &4380   &3.00   &15.2  &8.46   &5.52   &169    &37.0   &11.0\\
            MSUCOF-1-Ni\ce{Cl2}    &4200   &2.96   &15.2  &9.65   &5.66   &169    &37.2   &11.0\\
            MSUCOF-1-Pd\ce{Cl2}    &3630   &2.58   &16.8  &7.09   &5.14   &139    &36.9   &9.69\\
            MSUCOF-1-Pt\ce{Cl2}    &2890   &2.09   &16.8  &19.3   &8.43   &169    &41.4   &8.71\\
            \\
            MSUCOF-2              &4980   &6.59   &22.7  &3.04   &3.11   &1.5    &33.2   &18.9\\
            MSUCOF-2-Co\ce{Cl2}   &4380   &4.25   &18.8  &8.60   &4.72   &169    &36.7   &14.4\\
            MSUCOF-2-Cu\ce{Cl2}   &3910   &4.20   &21.1  &5.64   &3.99   &92     &34.5   &13.6\\
            MSUCOF-2-Fe\ce{Cl2}   &4440   &4.58   &18.7  &8.23   &4.54   &139    &35.6   &14.2\\
            MSUCOF-2-Mn\ce{Cl2}   &4420   &3.29   &18.7  &8.28   &4.59   &169    &36.1   &14.9\\
            MSUCOF-2-Ni\ce{Cl2}   &4310   &4.25   &18.9  &8.94   &4.64   &92     &35.8   &14.2\\
            MSUCOF-2-Pd\ce{Cl2}   &3480   &3.78   &21.2  &7.03   &4.44   &169    &36.2   &12.9\\
            MSUCOF-2-Pt\ce{Cl2}   &2890   &3.12   &21.1  &19.8   &6.77   &169    &38.7   &11.4\\
            \\
            MSUCOF-3               &5020   &1.90   &9.64  &3.85   &4.05   &-      &28.4   &6.35\\
            MSUCOF-3-CoCl          &3770   &0.94   &7.80  &16.2   &13.7   &139    &44.3   &5.22\\
            MSUCOF-3-CuCl          &3650   &0.91   &7.84  &10.5   &9.17   &139    &36.7   &4.38\\
            MSUCOF-3-FeCl          &3810   &0.95   &7.87  &16.2   &13.1   &113    &42.2   &5.10\\
            MSUCOF-3-MnCl          &3840   &0.95   &7.73  &15.9   &13.2   &139    &43.1   &5.20\\
            MSUCOF-3-NiCl          &3690   &0.93   &7.85  &17.2   &13.6   &139    &41.7   &4.95\\
            MSUCOF-3-PdCl          &2970   &0.75   &7.82  &13.1   &11.4   &139    &42.7   &4.20\\
            MSUCOF-3-PtCl          &2190   &0.55   &7.78  &26.8   &21.4   &75     &32.6   &2.29\\
        \hline
	\end{tabular}
	\caption{Tabulated properties of the three MSUCOF families showing the calculated: surface areas (\textit{\textbf{S$_\mathbf{A}$}}), pore volume (\textit{\textbf{V$_\mathbf{P}$}}), pore size (\textit{\textbf{P$_{\mathbf{size}}$}}), initial isosteric heat of adsorption at 1 bar (\textbf{Q$_{\mathbf{st}}^\mathbf{0}$}), isosteric heat of adsorption at 700 bar (\textbf{Q$_{\mathbf{st}}$}), optimal pressure (\textbf{\textit{P$_{\mathbf{opt}}$}}), volumetric working capacity, and gravimetric working capacity.}\label{summary}
\end{table}

\doublespacing 

\section{SUMMARY. }

The need for lightweight adsorbents that can operate under ambient conditions is critical for hydrogen storage in fuel cell vehicles. COFs have emerged as a promising candidate for this application, owing to their similar properties to MOFs, while being composed of fewer heavy elements. By incorporating transition metals into COFs and placing them in accessible areas within the pore, their interaction with gaseous components can be optimized. By carefully selecting linkers, researchers can generate COFs that are tailored for hydrogen storage applications. Increasing the number of binding sites per linker not only enhances the number of sites for \ce{H2} physisorption or chemisorption, but also enables overlapping interactions between these sites, resulting in a compounding gas-framework interaction. In this study, the authors have developed 24 new promising COFs using tri-topic linkers chelated with first-row transition metals and precious metals. MSUCOF-3-CoCl, MSUCOF-3-FeCl, MSUCOF-3-MnCl, MSUCOF-3-NiCl, MSUCOF-3-PdCl, and MSUCOF-1-Pt\ce{Cl2} have all surpassed DOE's 2025 40 g L$^{-1}$ working capacity target at ambient temperatures. Among these, MSUCOF-3-CoCl is particularly exceptional, achieving an impressive 55.7 absolute g L$^{-1}$ and 6.66 absolute wt \% uptakes, without the need for expensive Pd or Pt.

\section{ACKNOWLEDGMENTS} 

This work was supported in part by computational resources and services provided by the Institute for Cyber-Enabled Research at Michigan State University. We also thank Prof. Srimanta Pakhira for initial helpful discussions.

		\section{ASSOCIATED CONTENT}
		
		\textbf{Supporting Information Available}\\

		\section{AUTHOR INFORMATION}

		\textbf{Corresponding Authors}\\
		jmendoza@msu.edu
		
		\noindent 
		\textbf{Notes}\\
		The authors declare no competing financial interests.
		
		\bibliography{bibliography}

		\clearpage
		\newpage



\section*{Supplementary material}

# Supporting Information

## MultiBinding Sites United in Covalent-Organic Frameworks (MSUCOF) for $H_2$ Storage and Delivery at Room Temperature

Marcus Djokic[1], Jose L. Mendoza-Cortes[1*]

[1] Department of Chemical Engineering and Material Science, Michigan State University, East Lansing, MI, 48824, USA.

jmendoza@msu.edu

# Contents

# List of Tables

# List of Figures

# Economic Analysis of MSUCOFs

Table S1: Table breaking down retail costs of MSUCOF linkers. *note information regarding linker 3 could not be sourced via Sigma-Aldrich and has therefore been left blank

|  | Linkers | | | |
| --- | --- | --- | --- | --- |
|  | **1** | **2** | **3*** | **4** |
| Quantity (gram) | 1 | 1 | - | 1 |
| Cost/gram | $661 | $205 | - | $516 |
| sku | BCH00019-1G | 764957-1G | - | ANV00019-1G |
| CAS number | 153035-55-3 | 105598-27-4 | - |  |
| Mw (g/mol) | 495.7 | 384.3 | - | 552.0 |
| cost/kmol | $1333.48 | $533.47 | - | $934.82 |

The price of linkers (Table S1) and chelation transition metal (TM) chlorides (Table S2) are tabulated using prices from Sigma-Aldrich. Prices are shown in the default quantity, price per gram, and price per kmol of material. To be specific, the chosen linkers are as follows: methanetetrayltetrakis (benzene-4,1-diyl) tetraboronic acid (**1**), dipyrazino [2,3-f:2',3'-h] quinoxaline-2,3,6,7,10,11-hexaol (**2**), diquinoxalino [2,3-a:2',3'-c] phenazine-2,3,8,9,14,15-hexaol (**3**), and 5,5',5",5"'-methanetetrayl-tetrakis (pyrimidine-2-carbonitrile) (**4**). Please note that these prices listed in both tables are for the raw materials sold directly from the manufacturer and ignore whether or not they can be synthesized at a discount using other starting precursors.

Table S2: Table breaking down retail costs of MSUCOF chelation metal chlorides. This chart confirms that the more abundant first row TM (Co, Ni, Mn, Fe, and Cu) are significantly cheaper than the precious TM (Pt and Pd).

|  | Weight (gram) | Cost | Cost ($/gram) | sku | CAS number | Mw (g/mol) | Cost ($/kmol) |
| --- | --- | --- | --- | --- | --- | --- | --- |
| $PtCl_2$ | 5 | $405 | $81 | 206091-5G | 10025-65-7 | 266.0 | $304.52 |
| $PdCl_2$ | 100 | $7840 | $78.4 | 205885-100G | 7647-10-01 | 177.3 | $442.12 |
| $CoCl_2$ | 500 | $590 | $1.18 | 232696-500G | 7646-79-9 | 129.8 | $9.09 |
| $NiCl_2$ | 250 | $230 | $0.92 | 339350-250G | 7718-54-9 | 129.6 | $7.10 |
| $MnCl_2$ | 500 | $49.1 | $0.0982 | 416479-500G | 7773-01-05 | 125.8 | $0.78 |
| $FeCl_2$ | 250 | $783 | $3.132 | 372870-250G | 7758-94-3 | 126.8 | $24.71 |
| $CuCl_2$ | 500 | $76 | $0.152 | 8182470500 | 7447-39-4 | 134.5 | $1.13 |

Note that the performant MSUCOF-3-TMCl (consisting of linker 4) is more inexpensive than MSUCOF-1-TMCl2 (consisting of linkers 1 and 2) and potentially MSUCOF-2-TMCl$_2$ (consisting of linkers 1 and 3). Additionally, in comparison to the cost of the linkers alone, the first-row chelation metal chlorides are two to three orders of magnitude cheaper. Importantly, even the abundant Mn, Cu, Ni, and Co often outperform Pd and Pt in terms of the working capacities of the MSUCOF families, yet are also shown to be orders of magnitude cheaper. Altogether, TM chelation, even with first row TMs, is substantiated as an optimal technique to drastically improve hydrogen uptake with minimal added cost. Thus, economic focus should be put into driving down the cost of these linkers consequently making the COF more affordable.

# Theoretical Method for Surface Area and Pore Volume

The reported surface areas and pore volumes were calculated using Materials Studio. Specifically, a VdW Surface is created by intersecting the Van der Waal (VdW) radii of the framework and a probe solvent radius. By evaluating this throughout the pore, one can recreate a VdW surface and calculate its corresponding surface area. For these calculations, a solvent radius of 1.445 Å is used, based on the kinetic diameter of Hydrogen of 2.89 Å. An example VdW surface is overlaid on MSUCOF-1 and its metalated case in Figure S1.

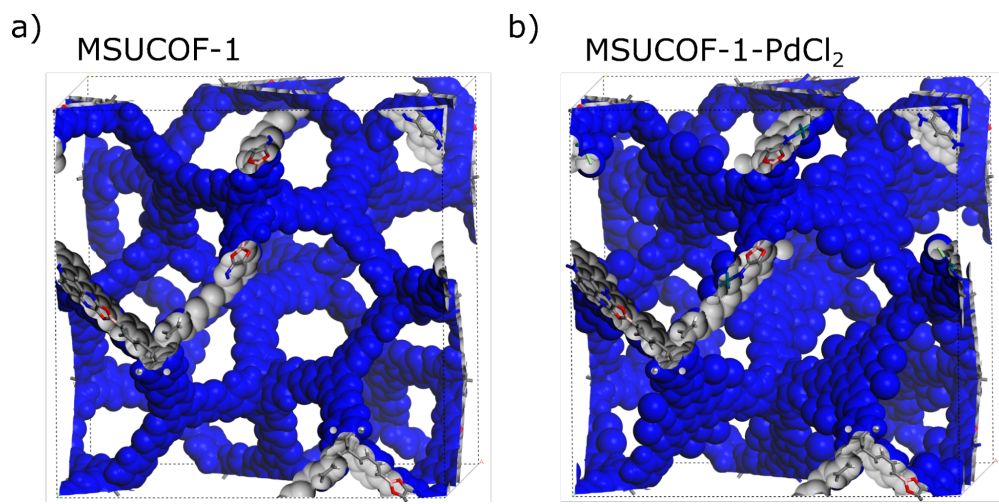

Figure S1: Van der Waal surface (in blue) projected onto a) MSUCOF-1 and b) metalated example MSUCOF-1-PdCl$_2$. This surface is used for pore volume and surface area calculations.

## Isotherm Plots

Calculations were performed to calculated absolute, net, and excess uptakes for each of the MSUCOFs. In the main manuscript, the absolute and net volumetric uptakes are shown as well as the absolute gravimetric plot. For completeness sake, the remaining excess volumetric, excess gravimetric, and net gravimetric plots are included below. Notably, the top performing MSUCOFs maintain exceptional net gravimetric uptakes of approximately 3% even at 700 bar. The optimal gavimetric pressures, as defined by maxima on the net gravimetric plots, tend to occur between 100-250 bar for the most exceptional COFs. This is in line with the results from the net volumetric uptake charts.

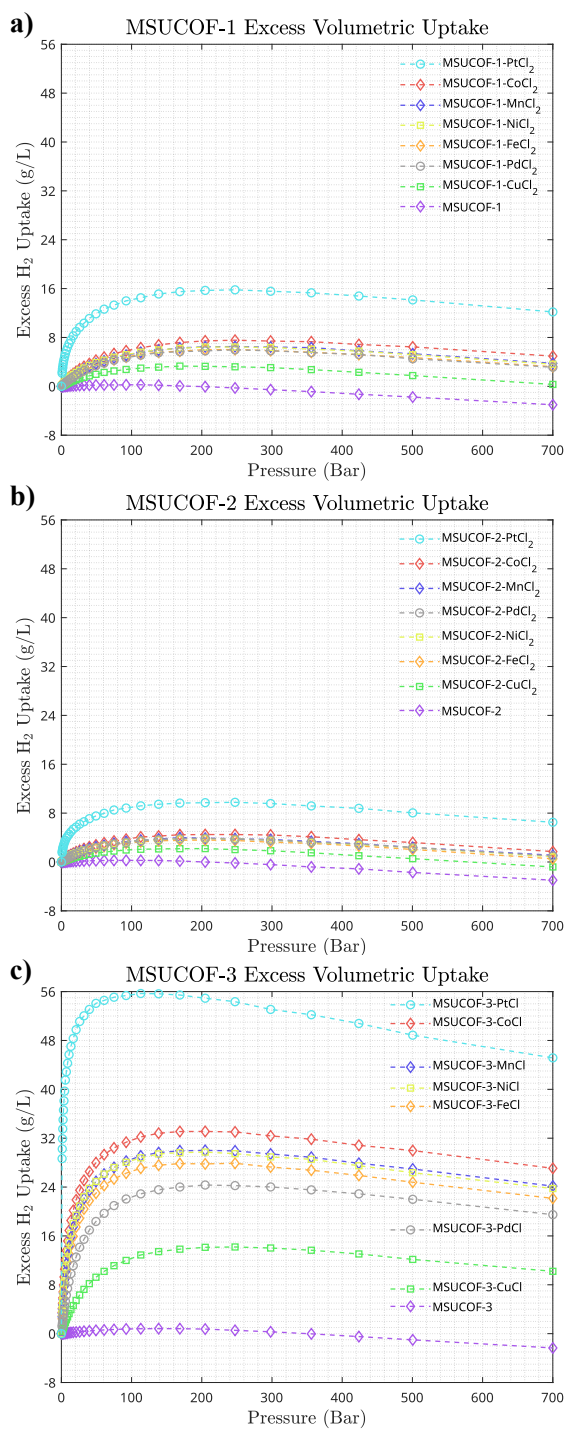

Figure S2: We show the high pressure $H_2$ excess volumetric isotherms, which compares uptakes with respect to $H_2$ loading into an equivalent pore volume, at 298 K for (a) MSUCOF-1-TMCl$_2$, (b) MSUCOF-2-TMCl$_2$, and (c) MSUCOF-3-TMCl.

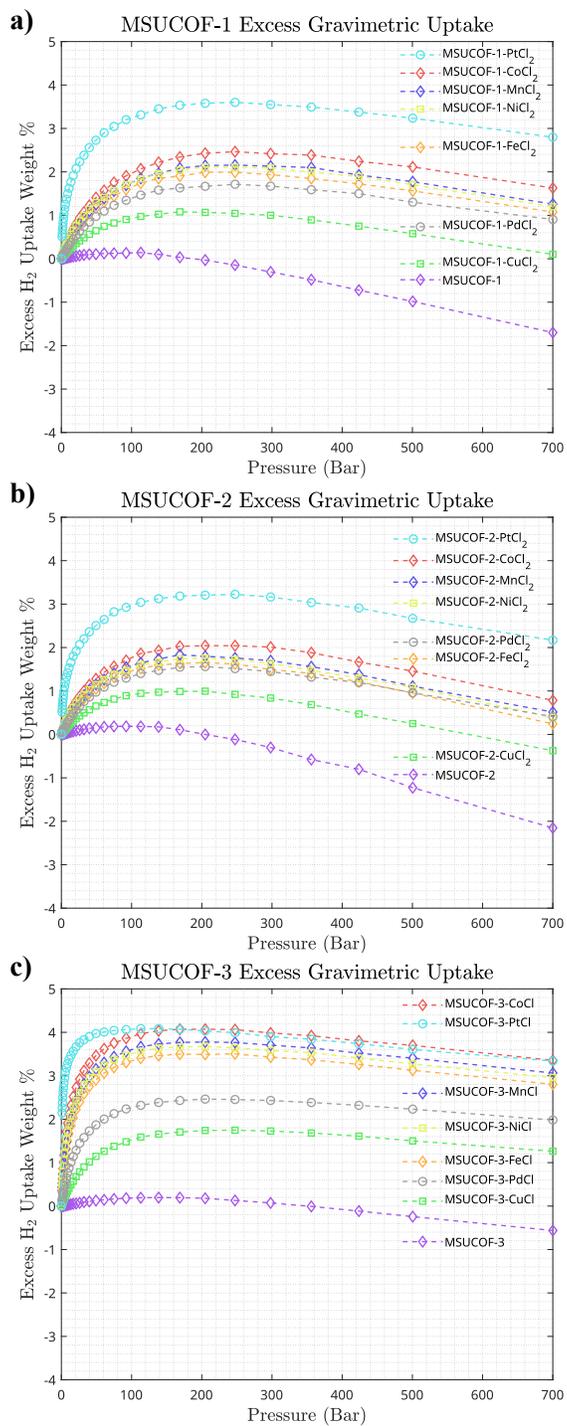

Figure S3: We show the high pressure $H_2$ excess gravimetric isotherms, which compares uptakes with respect to $H_2$ loading into an equivalent pore volume, at 298 K for (a) MSUCOF-1-TMCl$_2$, (b) MSUCOF-2-TMCl$_2$, and (c) MSUCOF-3-TMCl.

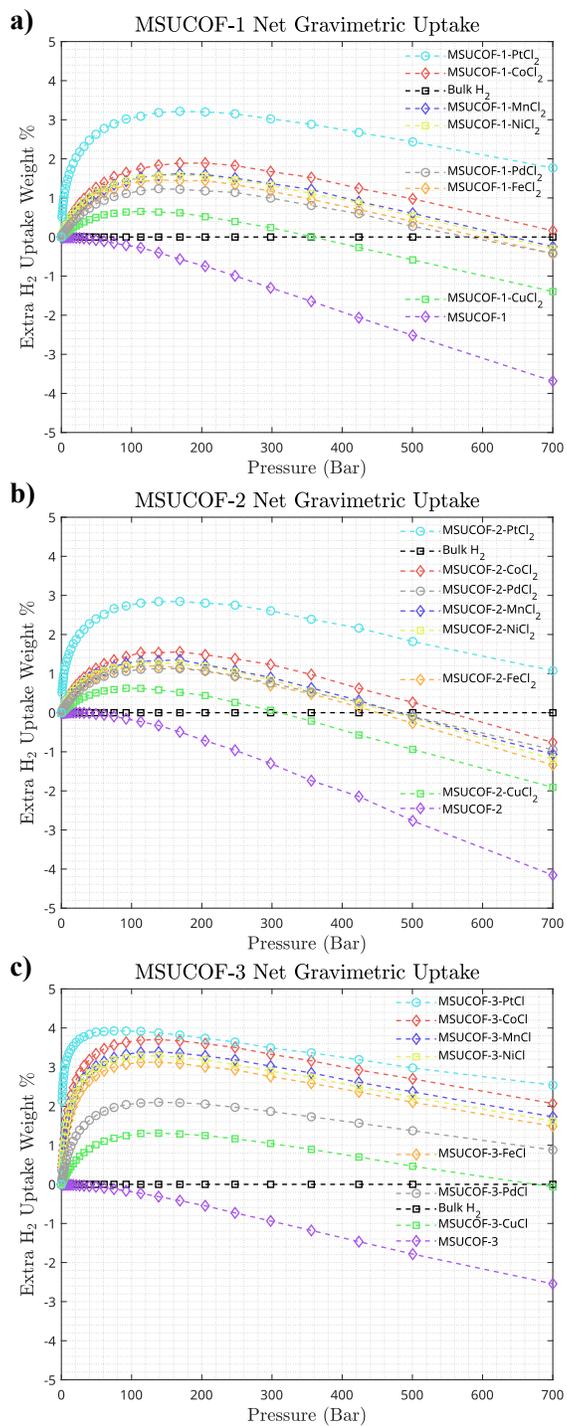

Figure S4: We show the high pressure $H_2$ net gravimetric isotherms, which compares uptakes with respect to $H_2$ loading in an empty unit cell of the same volume, at 298 K for (a) MSUCOF-1-TMCl$_2$, (b) MSUCOF-2-TMCl$_2$, and (c) MSUCOF-3-TMCl.

# Optimized Geometries

## MSUCOF-1

```
data_MSUCOF-1
_symmetry_space_group_name_H-M    'I-43D'
_symmetry_Int_Tables_number       220
_symmetry_cell_setting            cubic
loop_
_symmetry_equiv_pos_as_xyz
  x,y,z
  -x+1/2,-y,z+1/2
  -x,y+1/2,-z+1/2
  x+1/2,-y+1/2,-z
  z,x,y
  z+1/2,-x+1/2,-y
  -z+1/2,-x,y+1/2
  -z,x+1/2,-y+1/2
  y,z,x
  -y,z+1/2,-x+1/2
  y+1/2,-z+1/2,-x
  -y+1/2,-z,x+1/2
  y+1/4,x+1/4,z+1/4
  -y+1/4,-x+3/4,z+3/4
  y+3/4,-x+1/4,-z+3/4
  -y+3/4,x+3/4,-z+1/4
  x+1/4,z+1/4,y+1/4
  -x+3/4,z+3/4,-y+1/4
  -x+1/4,-z+3/4,y+3/4
  x+3/4,-z+1/4,-y+3/4
  z+1/4,y+1/4,x+1/4
  z+3/4,-y+1/4,-x+3/4
  -z+3/4,y+3/4,-x+1/4
  -z+1/4,-y+3/4,x+3/4
  x+1/2,y+1/2,z+1/2
  -x,-y+1/2,z
  -x+1/2,y,-z
  x,-y,-z+1/2
  z+1/2,x+1/2,y+1/2
  z,-x,-y+1/2
  -z,-x+1/2,y
  -z+1/2,x,-y
  y+1/2,z+1/2,x+1/2
  -y+1/2,z,-x
  y,-z,-x+1/2
  -y,-z+1/2,x
  y+3/4,x+3/4,z+3/4
  -y+3/4,-x+1/4,z+1/4
  y+1/4,-x+3/4,-z+1/4
  -y+1/4,x+1/4,-z+3/4
  x+3/4,z+3/4,y+3/4
  -x+1/4,z+1/4,-y+3/4
```

```
  -x+3/4,-z+1/4,y+1/4
  x+1/4,-z+3/4,-y+1/4
  z+3/4,y+3/4,x+3/4
  z+1/4,-y+3/4,-x+1/4
  -z+1/4,y+1/4,-x+3/4
  -z+3/4,-y+1/4,x+1/4
_cell_length_a                  44.3818
_cell_length_b                  44.3818
_cell_length_c                  44.3818
_cell_angle_alpha               90.0000
_cell_angle_beta                90.0000
_cell_angle_gamma               90.0000
loop_
_atom_site_label
_atom_site_type_symbol
_atom_site_fract_x
_atom_site_fract_y
_atom_site_fract_z
_atom_site_U_iso_or_equiv
_atom_site_adp_type
_atom_site_occupancy
O1    O    0.11672    0.24923    0.26434    0.00000    Uiso    1.00
N2    N    0.17328    0.19137    0.25958    0.00000    Uiso    1.00
C3    C    0.14166    0.23455    0.25261    0.00000    Uiso    1.00
C4    C    0.21781    0.18335    0.22645    0.00000    Uiso    1.00
O5    O    0.20284    0.29149    0.12931    0.00000    Uiso    1.00
N6    N    0.15826    0.24387    0.22787    0.00000    Uiso    1.00
C7    C    0.26778    0.14873    0.20882    0.00000    Uiso    1.00
B8    B    0.22877    0.28905    0.10906    0.00000    Uiso    1.00
C9    C    0.19174    0.20018    0.23477    0.00000    Uiso    1.00
H10   H    0.19143    0.33517    0.08908    0.00000    Uiso    1.00
C11   C    0.21265    0.33294    0.07478    0.00000    Uiso    1.00
C12   C    0.26008    0.30811    0.06410    0.00000    Uiso    1.00
C13   C    0.26475    0.32755    0.03978    0.00000    Uiso    1.00
C14   C    0.24348    0.34987    0.03288    0.00000    Uiso    1.00
H15   H    0.19965    0.37007    0.04464    0.00000    Uiso    1.00
H16   H    0.28597    0.32521    0.02550    0.00000    Uiso    1.00
C17   C    0.23403    0.31075    0.08180    0.00000    Uiso    1.00
C18   C    0.21728    0.35228    0.05035    0.00000    Uiso    1.00
H19   H    0.27749    0.29004    0.06960    0.00000    Uiso    1.00
C20   C    0.12500    0.00000    0.75000    0.00000    Uiso    1.00
```

# MSUCOF-2

```
data_MSUCOF-2
_symmetry_space_group_name_H-M    'I-43D'
_symmetry_Int_Tables_number       220
_symmetry_cell_setting            cubic
loop_
_symmetry_equiv_pos_as_xyz
  x,y,z
  -x+1/2,-y,z+1/2
  -x,y+1/2,-z+1/2
  x+1/2,-y+1/2,-z
```

```
    z,x,y
    z+1/2,-x+1/2,-y
    -z+1/2,-x,y+1/2
    -z,x+1/2,-y+1/2
    y,z,x
    -y,z+1/2,-x+1/2
    y+1/2,-z+1/2,-x
    -y+1/2,-z,x+1/2
    y+1/4,x+1/4,z+1/4
    -y+1/4,-x+3/4,z+3/4
    y+3/4,-x+1/4,-z+3/4
    -y+3/4,x+3/4,-z+1/4
    x+1/4,z+1/4,y+1/4
    -x+3/4,z+3/4,-y+1/4
    -x+1/4,-z+3/4,y+3/4
    x+3/4,-z+1/4,-y+3/4
    z+1/4,y+1/4,x+1/4
    z+3/4,-y+1/4,-x+3/4
    -z+3/4,y+3/4,-x+1/4
    -z+1/4,-y+3/4,x+3/4
    x+1/2,y+1/2,z+1/2
    -x,-y+1/2,z
    -x+1/2,y,-z
    x,-y,-z+1/2
    z+1/2,x+1/2,y+1/2
    z,-x,-y+1/2
    -z,-x+1/2,y
    -z+1/2,x,-y
    y+1/2,z+1/2,x+1/2
    -y+1/2,z,-x
    y,-z,-x+1/2
    -y,-z+1/2,x
    y+3/4,x+3/4,z+3/4
    -y+3/4,-x+1/4,z+1/4
    y+1/4,-x+3/4,-z+1/4
    -y+1/4,x+1/4,-z+3/4
    x+3/4,z+3/4,y+3/4
    -x+1/4,z+1/4,-y+3/4
    -x+3/4,-z+1/4,y+1/4
    x+1/4,-z+3/4,-y+1/4
    z+3/4,y+3/4,x+3/4
    z+1/4,-y+3/4,-x+1/4
    -z+1/4,y+1/4,-x+3/4
    -z+3/4,-y+1/4,x+1/4
_cell_length_a                    51.8991
_cell_length_b                    51.8991
_cell_length_c                    51.8991
_cell_angle_alpha                 90.0000
_cell_angle_beta                  90.0000
_cell_angle_gamma                 90.0000
loop_
_atom_site_label
_atom_site_type_symbol
_atom_site_fract_x
```

```
_atom_site_fract_y
_atom_site_fract_z
_atom_site_U_iso_or_equiv
_atom_site_adp_type
_atom_site_occupancy
C1     C     0.21333    0.28428    0.12054    0.00000   Uiso   1.00
C2     C     0.24302    0.24913    0.12810    0.00000   Uiso   1.00
H3     H     0.27427    0.30630    0.05572    0.00000   Uiso   1.00
C4     C     0.19831    0.27744    0.14175    0.00000   Uiso   1.00
H5     H     0.28227    0.33744    0.01937    0.00000   Uiso   1.00
H6     H     0.19722    0.33916    0.06734    0.00000   Uiso   1.00
C7     C     0.21632    0.33889    0.05650    0.00000   Uiso   1.00
C8     C     0.25879    0.32089    0.05027    0.00000   Uiso   1.00
C9     C     0.23545    0.32134    0.06375    0.00000   Uiso   1.00
C10    C     0.26317    0.33825    0.03017    0.00000   Uiso   1.00
C11    C     0.24440    0.35670    0.02304    0.00000   Uiso   1.00
C12    C     0.22048    0.35599    0.03613    0.00000   Uiso   1.00
H13    H     0.20425    0.36951    0.03010    0.00000   Uiso   1.00
C14    C     0.23520    0.27067    0.11404    0.00000   Uiso   1.00
B15    B     0.23060    0.30233    0.08663    0.00000   Uiso   1.00
N16    N     0.19199    0.24877    0.17732    0.00000   Uiso   1.00
C17    C     0.20586    0.25600    0.15636    0.00000   Uiso   1.00
O18    O     0.24774    0.28074    0.09274    0.00000   Uiso   1.00
C19    C     0.22811    0.24169    0.14940    0.00000   Uiso   1.00
N20    N     0.23484    0.22076    0.16357    0.00000   Uiso   1.00
O21    O     0.20858    0.30477    0.10414    0.00000   Uiso   1.00
C22    C     0.19888    0.22792    0.19166    0.00000   Uiso   1.00
C23    C     0.22078    0.21338    0.18446    0.00000   Uiso   1.00
H24    H     0.26097    0.23767    0.12272    0.00000   Uiso   1.00
H25    H     0.18035    0.28886    0.14720    0.00000   Uiso   1.00
C26    C     0.12500    0.00000    0.75000    0.00000   Uiso   1.00
```

# MSUCOF-3

```
data_MSUCOF-3
_symmetry_space_group_name_H-M    'I-43D'
_symmetry_Int_Tables_number       220
_symmetry_cell_setting            cubic
loop_
_symmetry_equiv_pos_as_xyz
  x,y,z
  -x+1/2,-y,z+1/2
  -x,y+1/2,-z+1/2
  x+1/2,-y+1/2,-z
  z,x,y
  z+1/2,-x+1/2,-y
  -z+1/2,-x,y+1/2
  -z,x+1/2,-y+1/2
  y,z,x
  -y,z+1/2,-x+1/2
  y+1/2,-z+1/2,-x
  -y+1/2,-z,x+1/2
  y+1/4,x+1/4,z+1/4
  -y+1/4,-x+3/4,z+3/4
```

```
  y+3/4,-x+1/4,-z+3/4
  -y+3/4,x+3/4,-z+1/4
  x+1/4,z+1/4,y+1/4
  -x+3/4,z+3/4,-y+1/4
  -x+1/4,-z+3/4,y+3/4
  x+3/4,-z+1/4,-y+3/4
  z+1/4,y+1/4,x+1/4
  z+3/4,-y+1/4,-x+3/4
  -z+3/4,y+3/4,-x+1/4
  -z+1/4,-y+3/4,x+3/4
  x+1/2,y+1/2,z+1/2
  -x,-y+1/2,z
  -x+1/2,y,-z
  x,-y,-z+1/2
  z+1/2,x+1/2,y+1/2
  z,-x,-y+1/2
  -z,-x+1/2,y
  -z+1/2,x,-y
  y+1/2,z+1/2,x+1/2
  -y+1/2,z,-x
  y,-z,-x+1/2
  -y,-z+1/2,x
  y+3/4,x+3/4,z+3/4
  -y+3/4,-x+1/4,z+1/4
  y+1/4,-x+3/4,-z+1/4
  -y+1/4,x+1/4,-z+3/4
  x+3/4,z+3/4,y+3/4
  -x+1/4,z+1/4,-y+3/4
  -x+3/4,-z+1/4,y+1/4
  x+1/4,-z+3/4,-y+1/4
  z+3/4,y+3/4,x+3/4
  z+1/4,-y+3/4,-x+1/4
  -z+1/4,y+1/4,-x+3/4
  -z+3/4,-y+1/4,x+1/4
_cell_length_a                 27.4078
_cell_length_b                 27.4078
_cell_length_c                 27.4078
_cell_angle_alpha              90.0000
_cell_angle_beta               90.0000
_cell_angle_gamma              90.0000
loop_
_atom_site_label
_atom_site_type_symbol
_atom_site_fract_x
_atom_site_fract_y
_atom_site_fract_z
_atom_site_U_iso_or_equiv
_atom_site_adp_type
_atom_site_occupancy
C1    C    0.33948    0.06876    0.19417    0.00000  Uiso   1.00
H2    H    0.36531    0.05751    0.16358    0.00000  Uiso   1.00
C3    C    0.34040    0.04365    0.23929    0.00000  Uiso   1.00
C4    C    0.30517    0.05696    0.27460    0.00000  Uiso   1.00
H5    H    0.30311    0.03639    0.31070    0.00000  Uiso   1.00
```

```
N6      N     0.27244   0.09512   0.26629   0.00000   Uiso   1.00
C7      C     0.27369   0.12105   0.22229   0.00000   Uiso   1.00
C8      C     0.23782   0.16465   0.21323   0.00000   Uiso   1.00
N9      N     0.30725   0.10744   0.18627   0.00000   Uiso   1.00
N10     N     0.19700   0.17173   0.24691   0.00000   Uiso   1.00
H11     H     0.19159   0.14613   0.27783   0.00000   Uiso   1.00
C12     C     0.37500   0.00000   0.25000   0.00000   Uiso   1.00
```

## Metalated MSUCOF-1 example: MSUCOF-1-CoCl$_2$

```
data_MSUCOF-1-CoCl2
_symmetry_space_group_name_H-M    'I-43D'
_symmetry_Int_Tables_number       220
_symmetry_cell_setting            cubic
loop_
_symmetry_equiv_pos_as_xyz
  x,y,z
  -x+1/2,-y,z+1/2
  -x,y+1/2,-z+1/2
  x+1/2,-y+1/2,-z
  z,x,y
  z+1/2,-x+1/2,-y
  -z+1/2,-x,y+1/2
  -z,x+1/2,-y+1/2
  y,z,x
  -y,z+1/2,-x+1/2
  y+1/2,-z+1/2,-x
  -y+1/2,-z,x+1/2
  y+1/4,x+1/4,z+1/4
  -y+1/4,-x+3/4,z+3/4
  y+3/4,-x+1/4,-z+3/4
  -y+3/4,x+3/4,-z+1/4
  x+1/4,z+1/4,y+1/4
  -x+3/4,z+3/4,-y+1/4
  -x+1/4,-z+3/4,y+3/4
  x+3/4,-z+1/4,-y+3/4
  z+1/4,y+1/4,x+1/4
  z+3/4,-y+1/4,-x+3/4
  -z+3/4,y+3/4,-x+1/4
  -z+1/4,-y+3/4,x+3/4
  x+1/2,y+1/2,z+1/2
  -x,-y+1/2,z
  -x+1/2,y,-z
  x,-y,-z+1/2
  z+1/2,x+1/2,y+1/2
  z,-x,-y+1/2
  -z,-x+1/2,y
  -z+1/2,x,-y
  y+1/2,z+1/2,x+1/2
  -y+1/2,z,-x
  y,-z,-x+1/2
  -y,-z+1/2,x
  y+3/4,x+3/4,z+3/4
  -y+3/4,-x+1/4,z+1/4
```

```
    y+1/4,-x+3/4,-z+1/4
    -y+1/4,x+1/4,-z+3/4
    x+3/4,z+3/4,y+3/4
    -x+1/4,z+1/4,-y+3/4
    -x+3/4,-z+1/4,y+1/4
    x+1/4,-z+3/4,-y+1/4
    z+3/4,y+3/4,x+3/4
    z+1/4,-y+3/4,-x+1/4
    -z+1/4,y+1/4,-x+3/4
    -z+3/4,-y+1/4,x+1/4
_cell_length_a                       44.3818
_cell_length_b                       44.3818
_cell_length_c                       44.3818
_cell_angle_alpha                    90.0000
_cell_angle_beta                     90.0000
_cell_angle_gamma                    90.0000
loop_
_atom_site_label
_atom_site_type_symbol
_atom_site_fract_x
_atom_site_fract_y
_atom_site_fract_z
_atom_site_U_iso_or_equiv
_atom_site_adp_type
_atom_site_occupancy
O1      O     0.11672    0.24923    0.26434    0.00000   Uiso   1.00
N2      N     0.17594    0.18930    0.26042    0.00000   Uiso   1.00
C3      C     0.14166    0.23455    0.25261    0.00000   Uiso   1.00
C4      C     0.21781    0.18335    0.22645    0.00000   Uiso   1.00
O5      O     0.20284    0.29149    0.12931    0.00000   Uiso   1.00
N6      N     0.15804    0.24495    0.22437    0.00000   Uiso   1.00
C7      C     0.26778    0.14873    0.20882    0.00000   Uiso   1.00
B8      B     0.22877    0.28905    0.10906    0.00000   Uiso   1.00
C9      C     0.19174    0.20018    0.23477    0.00000   Uiso   1.00
H10     H     0.19143    0.33517    0.08908    0.00000   Uiso   1.00
C11     C     0.21265    0.33294    0.07478    0.00000   Uiso   1.00
C12     C     0.26008    0.30811    0.06410    0.00000   Uiso   1.00
C13     C     0.26475    0.32755    0.03978    0.00000   Uiso   1.00
C14     C     0.24348    0.34987    0.03288    0.00000   Uiso   1.00
H15     H     0.19965    0.37007    0.04464    0.00000   Uiso   1.00
H16     H     0.28597    0.32521    0.02550    0.00000   Uiso   1.00
C17     C     0.23403    0.31075    0.08180    0.00000   Uiso   1.00
C18     C     0.21728    0.35228    0.05035    0.00000   Uiso   1.00
H19     H     0.27749    0.29004    0.06960    0.00000   Uiso   1.00
Co20    Co    0.40278    0.44474    0.52793    0.00000   Uiso   1.00
Cl21    Cl    0.36209    0.41686    0.52292    0.00000   Uiso   1.00
Cl22    Cl    0.40992    0.46364    0.57320    0.00000   Uiso   1.00
C23     C     0.12500   -0.00000    0.75000    0.00000   Uiso   1.00
```

# Metalated MSUCOF-2 example: MSUCOF-2-CoCl$_2$

```
data_MSUCOF-2-CoCl2
_symmetry_space_group_name_H-M       'I-43D'
_symmetry_Int_Tables_number          220
```

```
_symmetry_cell_setting          cubic
loop_
_symmetry_equiv_pos_as_xyz
  x,y,z
  -x+1/2,-y,z+1/2
  -x,y+1/2,-z+1/2
  x+1/2,-y+1/2,-z
  z,x,y
  z+1/2,-x+1/2,-y
  -z+1/2,-x,y+1/2
  -z,x+1/2,-y+1/2
  y,z,x
  -y,z+1/2,-x+1/2
  y+1/2,-z+1/2,-x
  -y+1/2,-z,x+1/2
  y+1/4,x+1/4,z+1/4
  -y+1/4,-x+3/4,z+3/4
  y+3/4,-x+1/4,-z+3/4
  -y+3/4,x+3/4,-z+1/4
  x+1/4,z+1/4,y+1/4
  -x+3/4,z+3/4,-y+1/4
  -x+1/4,-z+3/4,y+3/4
  x+3/4,-z+1/4,-y+3/4
  z+1/4,y+1/4,x+1/4
  z+3/4,-y+1/4,-x+3/4
  -z+3/4,y+3/4,-x+1/4
  -z+1/4,-y+3/4,x+3/4
  x+1/2,y+1/2,z+1/2
  -x,-y+1/2,z
  -x+1/2,y,-z
  x,-y,-z+1/2
  z+1/2,x+1/2,y+1/2
  z,-x,-y+1/2
  -z,-x+1/2,y
  -z+1/2,x,-y
  y+1/2,z+1/2,x+1/2
  -y+1/2,z,-x
  y,-z,-x+1/2
  -y,-z+1/2,x
  y+3/4,x+3/4,z+3/4
  -y+3/4,-x+1/4,z+1/4
  y+1/4,-x+3/4,-z+1/4
  -y+1/4,x+1/4,-z+3/4
  x+3/4,z+3/4,y+3/4
  -x+1/4,z+1/4,-y+3/4
  -x+3/4,-z+1/4,y+1/4
  x+1/4,-z+3/4,-y+1/4
  z+3/4,y+3/4,x+3/4
  z+1/4,-y+3/4,-x+1/4
  -z+1/4,y+1/4,-x+3/4
  -z+3/4,-y+1/4,x+1/4
_cell_length_a                  51.8991
_cell_length_b                  51.8991
_cell_length_c                  51.8991
```

```
_cell_angle_alpha                    90.0000
_cell_angle_beta                     90.0000
_cell_angle_gamma                    90.0000
loop_
_atom_site_label
_atom_site_type_symbol
_atom_site_fract_x
_atom_site_fract_y
_atom_site_fract_z
_atom_site_U_iso_or_equiv
_atom_site_adp_type
_atom_site_occupancy
Co1    Co    0.15934    0.26650    0.19490    0.00000    Uiso    1.00
Cl2    Cl    0.16817    0.30474    0.21093    0.00000    Uiso    1.00
C3     C     0.23520    0.27067    0.11404    0.00000    Uiso    1.00
H4     H     0.27427    0.30630    0.05573    0.00000    Uiso    1.00
C5     C     0.24302    0.24913    0.12810    0.00000    Uiso    1.00
C6     C     0.21333    0.28428    0.12054    0.00000    Uiso    1.00
C7     C     0.21632    0.33889    0.05650    0.00000    Uiso    1.00
C8     C     0.23545    0.32134    0.06375    0.00000    Uiso    1.00
C9     C     0.26317    0.33825    0.03017    0.00000    Uiso    1.00
Cl10   Cl    0.12250    0.26341    0.17417    0.00000    Uiso    1.00
C11    C     0.25879    0.32089    0.05027    0.00000    Uiso    1.00
H12    H     0.20431    0.36954    0.03004    0.00000    Uiso    1.00
B13    B     0.23060    0.30233    0.08663    0.00000    Uiso    1.00
N14    N     0.18873    0.24939    0.17831    0.00000    Uiso    1.00
O15    O     0.20858    0.30477    0.10414    0.00000    Uiso    1.00
C16    C     0.24440    0.35670    0.02304    0.00000    Uiso    1.00
C17    C     0.22048    0.35599    0.03613    0.00000    Uiso    1.00
C18    C     0.20586    0.25600    0.15636    0.00000    Uiso    1.00
H19    H     0.28229    0.33740    0.01940    0.00000    Uiso    1.00
C20    C     0.22811    0.24169    0.14940    0.00000    Uiso    1.00
C21    C     0.19831    0.27744    0.14175    0.00000    Uiso    1.00
O22    O     0.24774    0.28074    0.09274    0.00000    Uiso    1.00
H23    H     0.19721    0.33917    0.06733    0.00000    Uiso    1.00
N24    N     0.23834    0.21964    0.16457    0.00000    Uiso    1.00
C25    C     0.19888    0.22792    0.19166    0.00000    Uiso    1.00
C26    C     0.22078    0.21338    0.18446    0.00000    Uiso    1.00
H27    H     0.26097    0.23769    0.12270    0.00000    Uiso    1.00
H28    H     0.18035    0.28885    0.14721    0.00000    Uiso    1.00
C29    C     0.12500    0.00000    0.75000    0.00000    Uiso    1.00
```

## Metalated MSUCOF-3 example: MSUCOF-3-CoCl

```
data_MSUCOF-3-CoCl
_symmetry_space_group_name_H-M       'I-43D'
_symmetry_Int_Tables_number          220
_symmetry_cell_setting               cubic
loop_
_symmetry_equiv_pos_as_xyz
  x,y,z
  -x+1/2,-y,z+1/2
  -x,y+1/2,-z+1/2
  x+1/2,-y+1/2,-z
```

```
  z,x,y
  z+1/2,-x+1/2,-y
  -z+1/2,-x,y+1/2
  -z,x+1/2,-y+1/2
  y,z,x
  -y,z+1/2,-x+1/2
  y+1/2,-z+1/2,-x
  -y+1/2,-z,x+1/2
  y+1/4,x+1/4,z+1/4
  -y+1/4,-x+3/4,z+3/4
  y+3/4,-x+1/4,-z+3/4
  -y+3/4,x+3/4,-z+1/4
  x+1/4,z+1/4,y+1/4
  -x+3/4,z+3/4,-y+1/4
  -x+1/4,-z+3/4,y+3/4
  x+3/4,-z+1/4,-y+3/4
  z+1/4,y+1/4,x+1/4
  z+3/4,-y+1/4,-x+3/4
  -z+3/4,y+3/4,-x+1/4
  -z+1/4,-y+3/4,x+3/4
  x+1/2,y+1/2,z+1/2
  -x,-y+1/2,z
  -x+1/2,y,-z
  x,-y,-z+1/2
  z+1/2,x+1/2,y+1/2
  z,-x,-y+1/2
  -z,-x+1/2,y
  -z+1/2,x,-y
  y+1/2,z+1/2,x+1/2
  -y+1/2,z,-x
  y,-z,-x+1/2
  -y,-z+1/2,x
  y+3/4,x+3/4,z+3/4
  -y+3/4,-x+1/4,z+1/4
  y+1/4,-x+3/4,-z+1/4
  -y+1/4,x+1/4,-z+3/4
  x+3/4,z+3/4,y+3/4
  -x+1/4,z+1/4,-y+3/4
  -x+3/4,-z+1/4,y+1/4
  x+1/4,-z+3/4,-y+1/4
  z+3/4,y+3/4,x+3/4
  z+1/4,-y+3/4,-x+1/4
  -z+1/4,y+1/4,-x+3/4
  -z+3/4,-y+1/4,x+1/4
_cell_length_a                  27.4078
_cell_length_b                  27.4078
_cell_length_c                  27.4078
_cell_angle_alpha               90.0000
_cell_angle_beta                90.0000
_cell_angle_gamma               90.0000
loop_
_atom_site_label
_atom_site_type_symbol
_atom_site_fract_x
```

```
_atom_site_fract_y
_atom_site_fract_z
_atom_site_U_iso_or_equiv
_atom_site_adp_type
_atom_site_occupancy
C1    C    0.33948   0.06876   0.19417   0.00000   Uiso   1.00
H2    H    0.36531   0.05751   0.16358   0.00000   Uiso   1.00
C3    C    0.34040   0.04365   0.23929   0.00000   Uiso   1.00
C4    C    0.30517   0.05696   0.27460   0.00000   Uiso   1.00
H5    H    0.30311   0.03639   0.31070   0.00000   Uiso   1.00
N6    N    0.25870   0.09695   0.26891   0.00000   Uiso   1.00
C7    C    0.27369   0.12105   0.22229   0.00000   Uiso   1.00
C8    C    0.23782   0.16465   0.21323   0.00000   Uiso   1.00
N9    N    0.31171   0.12243   0.18278   0.00000   Uiso   1.00
N10   N    0.19201   0.16865   0.23796   0.00000   Uiso   1.00
Co11  Co   0.43809   0.54812   0.38210   0.00000   Uiso   1.00
Cl12  Cl   0.43326   0.61926   0.33879   0.00000   Uiso   1.00
C13   C    0.37500   0.00000   0.25000   0.00000   Uiso   1.00
```

# TOC Graphic

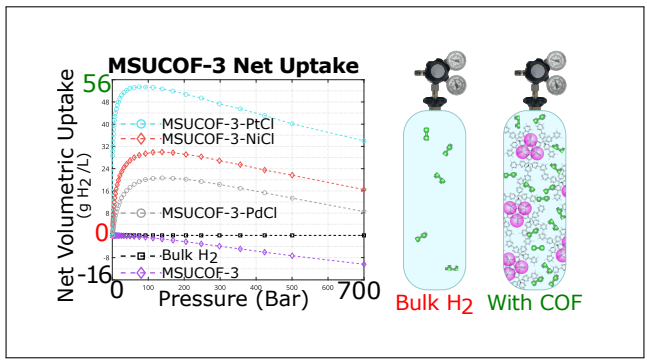


	
\end{document}